\documentclass{article}
\usepackage[margin=1in]{geometry}
\usepackage{booktabs}
\usepackage{graphicx} 
\usepackage{amsmath}
\usepackage{amsfonts}
\usepackage[numbers]{natbib}
\usepackage{float}
\usepackage{xcolor}
\usepackage{rotating}

\title{Targeted maximum likelihood estimation for longitudinal two-stage designs with outcome subsampling}
\author{
Kirsten E. Landsiedel$^{1,*}$, Maya L. Petersen$^{1}$, Mark J. van der Laan$^{1}$ \\[0.5em]
$^{1}$Division of Biostatistics, School of Public Health, University of California, Berkeley, CA 94720, USA \\[0.3em]
$^{*}$Corresponding author: \texttt{kirsten\_landsiedel@berkeley.edu}}
\date{}

\begin{document}

\maketitle

\section*{Abstract}

We consider efficient estimation of general causal parameters in longitudinal 
two-stage designs with outcome subsampling, motivated by resampling 
designs arising in longitudinal survival studies of HIV-related mortality 
in resource-limited settings. In these studies, a substantial proportion 
of participants become lost to follow-up before either death is recorded or right censoring time is reached; resampling designs address this by retrospectively tracing a 
subset of lost individuals to ascertain their outcomes. Standard analyses often 
rely on inverse-probability-weighted Kaplan-Meier (wKM) estimators that 
discard the rich longitudinal covariate history available in these studies 
and suffer from substantial efficiency losses. A key observation of this paper is that resampling designs are an instance of a broader class: two-stage designs with outcome subsampling, 
in which a first stage collects some data on all participants and a second 
stage collects outcome information on a selected subset. This connection 
motivates two novel estimators applicable to the general class. First, 
drawing on inverse probability of censoring weighted targeted maximum likelihood estimation (IPCW-TMLE) for two-stage designs, we develop its longitudinal extension, IPCW longitudinal TMLE (IPCW-LTMLE) and show that estimating and targeting the known
second-stage sampling weights yields variance reductions of up to 36\% 
over the use of known sampling probabilities. Second, observing that 
inverse weighting sacrifices efficiency, we propose an LTMLE that 
incorporates the second-stage sampling indicator directly as an 
intervention node in the sequential regression framework, returning to 
plug-in estimation principles and avoiding inverse weighting entirely. 
Using the resampling setting as our concrete example, simulations across 
$N \in \{500, 1000, 3000\}$ show that LTMLE achieves up to 73\% lower 
variance than wKM with known sampling weights, with reductions of 
30--50\% common across settings, while IPCW-LTMLE achieves consistent gains of 
20--35\%. We further demonstrate that cross-fitted variance estimation is 
essential for valid inference: standard variance estimators yield 
confidence interval coverage as low as 76\%, while our proposed 
cross-fitted variants consistently restore coverage to nominal levels. In 
settings with bivariate censoring, as in resampling designs, fully 
efficient closed-form estimators do not exist in general; our proposed 
estimators represent highly efficient closed-form alternatives that 
substantially expand the methodological toolkit for two-stage designs 
with outcome subsampling.

\bigskip
\noindent \textbf{Keywords:} two-stage designs; outcome subsampling; targeted maximum likelihood estimation; cross-fitting; double sampling

 \newpage

\section{Introduction}
\label{sec:intro}

In longitudinal survival studies conducted in resource-limited settings, 
patients are monitored through repeated clinic visits, but a substantial 
proportion become lost to follow-up (LTFU) before a failure event or a censoring event is recorded. Because LTFU patients tend to be sicker and 
experience higher mortality than those retained in care, survival 
estimates based solely on observed clinical data are typically biased 
upward \citep{geng2013failure, holmes2018estimated}. Resampling designs 
address this by retrospectively selecting a subset of LTFU individuals 
for active tracing to ascertain their vital status.

The dominant approach for survival estimation in resampling designs is 
the inverse-probability-weighted Kaplan-Meier (wKM) estimator 
\citep{geng2012causal, geng2015estimation, holmes2018estimated, 
yiannoutsos2008sampling}, which assign a weight of one to participants with 
outcomes known from clinical records, a weight inverse to the known (or empirical) resampling-probability to participants chosen for resampling, and a weight of zero to participants 
whose outcomes remain unknown after tracing. Though unbiased, wKM 
estimators ignore the longitudinal covariate history available in these studies, 
even when that history is highly predictive of survival. 

In formalizing the data structure common in resampling designs, one can see that resampling designs are actually an
instance of a broader and well-known class: \emph{two-stage designs}. In two-stage designs, a first stage collects 
data $V$ on all study participants, and a second stage collects 
additional information on a selected subset. Let $\Delta \in \{0,1\}$ denote the stage-two sampling indicator, where 
$\Delta = 1$ indicates selection for additional data collection, and let 
$X$ denote the full data structure that would be observed if all variables 
were measured on all participants. We note that this definition of the full data $X$ is flexible: 
in more complex settings, it is sometimes convenient to define $X$ as a 
data structure that itself contains censoring or missingness, with the 
two-stage sampling layered on top. We adopt precisely this convention in 
our motivating resampling application, discussed in detail 
in Section~\ref{subsec:notation}. The observed data are then $O = (V, \Delta, \Delta X)$, 
so that $X$ is only observed when $\Delta = 1$ and $V \subset X$. This 
framework accommodates a variety of designs depending on which components 
of $X$ are observed only at stage two; we give three illustrative 
examples below.

\paragraph{Example 1 (Covariate subsampling)}
Let the full data be $X = (W_1, W_2, A, Y)$, where $W_1$ and $W_2$ are 
covariates, $A$ is treatment, and $Y$ is the outcome. Suppose that $W_1$, 
$A$, and $Y$ are observed for all participants, while $W_2$ is only 
observed at stage two. Then the observed data are
\[
O = (W_1, A, Y, \Delta, \Delta W_2).
\]
This design is particularly useful when some covariates are expensive to 
measure and resource constraints prevent their assessment in all 
participants \citep{wang2009causal, breslow1997maximum, 
williamson2026assessing, qiu2026efficient}.

\paragraph{Example 2 (Treatment subsampling)}
Let the full data be $X = (W, A, Y)$, where $W$ is the sole covariate. 
Suppose that $W$ and $Y$ are observed for all participants, while $A$ is 
only observed at stage two. Then the observed data are
\[
O = (W, Y, \Delta, \Delta A).
\]
This design is most similar in practice to a case-cohort or test-negative 
design, where the measurement of treatment depends on the outcome, as in 
studies of influenza vaccine effectiveness \citep{foppa2013case}.

\paragraph{Example 3 (Outcome subsampling)}
Let the full data be $X = (W, A, Y)$. Suppose that $W$ and $A$ are 
observed for all participants, while $Y$ is only observed at stage two. 
Then the observed data are
\[
O = (W, A, \Delta, \Delta Y).
\]

Our resampling design application corresponds precisely to Example 3; where outcome subsampling occurs in stage two. In this longitudinal 
survival setting, we define the full data as $X = (W, \tau, \bar{L}(\tau), 
\bar{Y}(\tau))$ which includes baseline covariates $W$, right censoring by administrative end of study $\tau$, time-varying covariate history $\bar{L}(\tau)$ through end of study, and the 
survival outcome process $\bar{Y}(\tau)$ through end of study. We use the overbar notation to denote the history of a random variable through a given time (e.g. $\bar{L}(t)=L(1),L(2),...,L(t)$). The stage-one data 
$V = (W, \tau, \bar{L}(\tau))$ are observed for all participants, while 
the outcome process $\bar{Y}(\tau)$ is ascertained only for participants 
with $\Delta = 1$ (those whose vital status is known either directly 
from clinical records or through active tracing in stage two). To our knowledge, the 
connection between resampling designs and the broader class of two-stage 
designs with outcome subsampling has not previously been carefully formalized in 
the literature. Establishing it allows us to draw on the two-stage design 
literature to motivate and construct more efficient estimators and 
reveals that the estimators we develop are useful well beyond the 
resampling setting itself.

Within the two-stage design literature, \citet{rose2011targeted} proposed 
inverse probability of censoring weighted targeted maximum likelihood estimation (IPCW-TMLE) for point-treatment settings that applies a full-data TMLE 
among second-stage participants, inversely weighted by the stage-two 
sampling probabilities. Recognizing that we can view the full data in the resampling 
setting as a longitudinal, right-censored survival data structure, we 
extend this framework to develop inverse probability of censoring weighted \textit{longitudinal} targeted maximum likelihood estimation (IPCW-LTMLE) for longitudinal two-stage 
designs, accommodating time-varying covariates and
censoring as needed. We further show that even when sampling probabilities are known by design, as is typical in two-stage studies, estimating these probabilities can substantially reduce variance by incorporating information from all participants; targeting the estimated probabilities via an additional fluctuation step yields further efficiency gains by enforcing a mean-zero influence curve condition.

Inverse probability weighting is known to sacrifice efficiency relative 
to plug-in estimation, and targeting the sampling weights only partially 
compensates for this loss. We thus propose an alternative procedure which incorporates the
second stage sampling mechanism directly into the sequential regression 
framework of LTMLE, avoiding inverse weighting entirely. We show 
that this is possible by introducing $\Delta$ as an additional 
intervention node in the observed data structure, so that the LTMLE 
estimates a counterfactual outcome under a joint intervention that prevents censoring and sets $\Delta = 1$ 
for all participants. This novel LTMLE implementation returns to plug-in 
estimation principles, leverages the full longitudinal covariate history 
of all participants regardless of their stage-two sampling status, and generally 
achieves some of the lowest empirical variance among all estimators considered in 
our simulations.

Both estimators are general to causal parameters expressible as marginal 
means in longitudinal two-stage designs with outcome subsampling. We 
develop and present them in the notation of our motivating resampling 
application, where the data structure introduces additional complexity 
from censoring (beyond the LTFU) and deterministic outcomes; 
resampling-specific implementation choices are called out explicitly 
where they arise. Additionally, we develop cross-fitted variance 
estimators for both proposed estimators. Standard influence-curve-based 
variance estimation yields confidence interval coverage as low as 76\% 
when flexible nuisance estimation is used; cross-fitted variance 
estimation restores coverage to near-nominal levels and is, we argue, 
an essential practice for applied use of LTMLE with data-adaptive 
nuisance estimation.

The remainder of this paper is organized as follows. In 
Section~\ref{sec:methods}, we present causal target parameters and 
identification results for longitudinal two-stage designs with outcome 
subsampling, using the resampling setting as a running example. In 
Section~\ref{subsec:estimation}, we review LTMLE and sequential 
regression before introducing our two proposed estimators. In 
Section~\ref{sec:cross}, we develop cross-fitted variance estimators for 
both proposed estimators. In Section~\ref{sec:sims}, we evaluate 
finite-sample performance via simulations in the resampling setting 
across sample sizes $N \in \{500, 1000, 3000\}$. Finally, in 
Section~\ref{sec:disc}, we discuss implications for estimation in 
two-stage designs with outcome subsampling more broadly and propose 
directions for future work.

\subsection{Literature Review}
\label{subsec:lit}

Two-stage designs have a long history in statistics and epidemiology
\citep{neyman1938contribution, cochran1977sampling, 
breslow1983multiplicative}. The majority of the two-stage design 
literature focuses on settings where the second stage collects covariate 
or exposure information, with the outcome observed for all participants 
at stage one \citep{williamson2026assessing, breslow1997maximum, 
breslow2003large, tao2020optimal}. \citet{robins1994estimation} developed 
the foundational missing data framework for two-stage designs, proposing 
inverse probability weighted estimating equation-based estimators; 
\citet{vanderlaan2003unified} provides a comprehensive treatment of 
efficient estimation for general missing and censored data structures, 
including those that arise in two-stage designs. Two-stage designs with 
outcome subsampling arise across a range of applied settings. The most 
prominent examples in survival analysis come from studies of the HIV 
care cascade in resource-limited settings, where high rates of loss to 
follow-up preclude complete measurement of the outcome 
\citep{geng2013failure, geng2015estimation, geng2012causal, 
holmes2018estimated, yiannoutsos2008sampling, an2015choosing, 
bakoyannis2020semiparametric}. Outcome subsampling also appears in 
randomized experiments where outcomes are missing due to survey 
nonresponse \citep{coppock2017combining} and in electronic health 
record studies where long-term outcomes are unavailable for a subset 
of patients \citep{sun2025estimating,barnatchez2025efficient}. These latter settings, however, 
focus primarily on estimation of average treatment effects from 
point-treatment data structures. The longitudinal and survival settings
with time-varying covariates, administrative censoring, and complex 
covariate histories that accumulate over follow-up are comparatively 
less developed.

As described above, the dominant approach for survival estimation in 
resampling designs is the weighted Kaplan-Meier (wKM) estimator, with inverse weighting by 
the stage-two sampling probabilities a common choice across two-stage 
designs of all types. \citet{frangakis2001addressing} proposed an 
alternative approach constructing survival curves by estimating the 
average hazard across dropout and non-dropout groups, though 
\citet{robins2001discussion} showed this to be equivalent to an 
inefficient IPW estimator. \citet{li2008non} extended Nelson-Aalen 
estimation to this setting, constructing survival estimators that borrow 
strength across stages to reduce finite-sample bias. 
\citet{bakoyannis2020semiparametric} proposed a sieve IPW estimator for 
competing risks under double-sampling designs, though this targets a 
conditional regression parameter rather than a marginal causal survival 
functional. More recently, \citet{levis2022double} derived efficient 
influence functions and one-step estimators for the causal average 
treatment effect under double sampling; \citet{frangakis2015deductive} 
and \citet{qian2020deductive} proposed a deductive approach using 
numerical Gateaux derivatives to avoid closed-form efficient influence 
function derivation, extended by the latter to resampling settings with 
time-varying follow-up, though at the cost of analytical tractability 
and transparency about conditions for robustness, efficiency, and 
boundedness. Of note, without second-stage data, inverse probability 
weighting can theoretically correct for informative loss to follow-up; 
however, this requires sequential ignorability of missingness given 
measured history. With resampled data, this strong assumption is no 
longer needed \citep{geng2012causal}.

The most directly relevant prior methodological work is 
\citet{rose2011targeted}, where a general IPCW-TMLE estimator was 
proposed for two-stage designs. Although \citet{rose2011targeted} 
discussed the possibility of additionally targeting the stage-two 
sampling mechanism to further reduce variance, this step was not 
implemented in their simulations or in the accompanying 
\texttt{twoStageDesignTMLE} R package \citep{gruber2025twostagedesigntmle}. 
Building on this, \citet{qiu2026efficient} implement a targeted IPCW-TMLE 
and propose a new class of TMLE-based estimators for two-stage designs; 
however, this new class of TMLE estimators relies on an assumption that 
stage-two information is structurally independent of stage-one data 
beyond the sampling weights, which is violated in our longitudinal 
resampling setting where the amount of data collected in stage two 
depends directly on a stage-one variable. Two-phase designs have also been used for measurement error correction 
in large observational databases \citep{lotspeich2022efficient}. Most 
relevant to our work, \citet{barnatchez2025efficient} develop 
semi-parametric efficient one-step estimators for the average treatment 
effect in a cross-sectional setting where both outcome and treatment are 
subject to measurement error and true values are recovered via two-phase 
sampling. They present two asymptotically equivalent approaches to 
constructing efficient estimators and demonstrate attractive finite-sample 
properties of inverse-weighted full-data estimators in this setting. Our 
IPCW-LTMLE is very similar in spirit to their Approach 2 estimator, and our 
targeted IPCW-LTMLE parallels their Approach 2 estimator with empirical 
efficiency maximization. In this sense, our work can be viewed as a 
longitudinal extension of theirs; a key distinction is that in our 
resampling setting, bivariate censoring by $(\tau, \Delta)$ precludes 
analytical derivation of the efficient influence curve, whereas the 
cross-sectional average treatment effect (ATE) setting of \citet{barnatchez2025efficient} admits a 
closed-form efficient influence curve directly.

\section{Methods}
\label{sec:methods}

The estimators proposed in this paper target general causal parameters 
expressible as marginal means in longitudinal two-stage designs with 
outcome subsampling, as defined by the observed data structure 
$O = (V, \Delta, \Delta X)$ introduced in Section~\ref{sec:intro}. 
We organize this section as follows. In 
Section~\ref{subsec:identification}, we present identification results 
at the level of the general class, using general longitudinal notation 
with resampling-specific illustrations provided in accompanying remarks. In Section~\ref{subsec:notation}, 
we introduce notation for our motivating resampling application, which 
serves as the running example throughout.
In Section~\ref{subsec:estimation}, we develop two estimators in the 
notation of the resampling setting; resampling-specific implementation 
choices are called out explicitly where they arise. The IPCW-LTMLE 
applies to general longitudinal two-stage designs, including those with 
covariate or exposure subsampling; the LTMLE applies specifically to 
two-stage designs with outcome subsampling, of which resampling is our 
motivating example.

\subsection{Causal target parameter and identification}
\label{subsec:identification}
We present identification results for general causal parameters that can be 
expressed as marginal means under progressively complex longitudinal data 
structures, building from the complete data setting through to the complex setting that arises in our resampling motivating example. For each data structure, we state a general identification result 
and accompany it with a note describing the specific form it takes in our resampling application as a running example. While the 
framework can additionally handle treatments $A(t)$, we restrict 
attention here to censoring and sampling mechanisms; extension to settings with 
time-varying treatment is the subject of ongoing work.

\subsubsection{No missingness}

Consider first the simplified setting in which all participants are followed for $K$ total time points. The observed data can be written in longitudinal form as
\[
O = (L(0), L(1), Y(1), L(2), Y(2), \ldots, L(K), Y(K)),
\]
where $Y(t)$ denotes the outcome measured at time $t$ and $L(t)$ denotes 
time-varying covariates. We follow patients for $K$ total time points but may choose to evaluate the target parameter at any time of interest $t_0 \leq K$; the observed data structure $O$ is the same for all choices of $t_0$, while the estimand changes with $t_0$. The general causal target parameter is a marginal 
mean of the outcome at some time of interest for estimation $t_0$ 
\[
\Psi(t_0) = E[Y(t_0)],
\]
which in this setting with no censoring or missingness is equal to its 
statistical counterpart directly and can be estimated consistently by the 
empirical mean $\frac{1}{n}\sum_{i=1}^n Y_i(t_0)$. 

\medskip
\noindent\textit{In our resampling application,
$Y(t) = \mathbb{I}(T > t)$ where $T$ denotes time of death yields 
$\Psi(t_0) = E[I(T > t_0)]$, the probability of surviving past $t_0$.}

\subsubsection{Right censoring}
\label{subsubsec:right_censoring}

We now introduce right censoring, so that the outcome at $t_0$ may not 
be observed for all participants; again we assume $K$ total time points 
of data are collected. The observed data can be written in longitudinal 
form as
\[
O = (L(0), C(1), L(1), Y(1), C(2), L(2), Y(2), \ldots, C(K), L(K), Y(K)),
\]
where $C(t)$ indicates whether a participant has remained uncensored at 
time $t$. After censoring, subsequent $L(t)$ and $Y(t)$ nodes are 
deterministically set to their last measured values. The general causal 
target parameter is a marginal mean of the outcome under an intervention to prevent 
censoring through $t_0$, that is, to set $\bar{C}(t_0) = 1$:
\[
\Psi(t_0) = E\bigl[Y(t_0)^{\bar{C}(t_0)=1}\bigr].
\]
Under sequential randomization of the censoring mechanism,
$Y(t_0)^{\bar{C}(t_0)=1} \perp C(t) \mid \bar{Y}(t-1), \bar{L}(t-1), 
\bar{C}(t-1) = 1$ for $t = 1, \ldots, t_0$, this is identified by the 
sequential regression representation of the longitudinal g-computation 
formula
\[
\Psi(t_0) = E\Big[E\Big[\cdots 
E\Big[E\bigl[Y(t_0) \mid \bar{L}(t_0), \bar{C}(t_0)=1, 
\bar{Y}(t_0-1)\bigr] 
\mid \bar{L}(t_0-1), \bar{C}(t_0-1)=1, 
\bar{Y}(t_0-2)\Big] \cdots \mid L(0)\Big]\Big].
\]
We discuss estimation of such quantities in Section~\ref{subsec:ltmle_over}.

\medskip
\noindent\textit{In our resampling application,
$Y(t) = \mathbb{I}(T > t)$, and 
the observed outcome at $t_0$ is the last observed value of $Y(t_0)$, 
given by $Y = \mathbb{I}(T > t_0,\, \bar{C}(t_0) = 1)$. This yields target parameter
$\Psi(t_0) = E\bigl[Y^{\bar{C}(t_0)=1}\bigr] = P(T > t_0)$, the 
counterfactual probability of surviving past $t_0$ had censoring been prevented through $t_0$.}

\subsubsection{Right censoring and outcome subsampling}
\label{subsubsec:right_censoring_ltfu}

We now consider an observed data structure arising in longitudinal 
two-stage designs with outcome subsampling and right censoring, where 
participants may additionally have unobserved outcomes after the first 
stage of data collection. As described previously, $\Delta$ indicates 
whether a participant's outcome is observed after second-stage data 
collection. The observed data can be written in longitudinal form as
\[
O = \bigl(L(0), C(1), L(1), C(2), \ldots, C(K), L(K), \Delta, \Delta Y(t_0)\bigr),
\]
where $Y(t_0)$ is the 
terminal outcome defined at $t_0$, placed after $\Delta$ to reflect that 
outcome ascertainment occurs after the second stage. More concretely, while the target timepoint $t_0$ may be strictly less 
than $K$, the observed data structure running through $L(K)$ before 
$\Delta$ and $\Delta Y(t_0)$ reflects the natural ordering of the study. 
Second-stage outcome ascertainment occurs at each participant's 
end of follow-up, so the full stage-one data through $K$ 
must be complete before second-stage sampling is finalized. In later sections we discuss the implications of this data structure on estimation procedures. Unlike the 
right-censoring setting where $Y(t)$ is a process evaluated at each $t$, 
here $Y(t_0)$ is only a terminal node; the observed data structure $O$ is the same 
for all choices of $t_0 \leq K$, while $Y(t_0)$ and the estimand change 
with $t_0$. The general causal target parameter is a marginal mean under 
a joint intervention to prevent censoring through $t_0$ and to set 
$\Delta = 1$:
\[
\Psi(t_0) = E\bigl[Y(t_0)^{\Delta=1,\, \bar{C}(t_0)=1}\bigr].
\]
Identification proceeds under sequential randomization of the censoring 
mechanism, as in Section~\ref{subsubsec:right_censoring}, together with 
randomization of the second-stage sampling mechanism:
\[
Y(t_0)^{\Delta=1,\,\bar{C}(t_0)=1} \perp \Delta \mid \bar{L}(K), 
\bar{C}(K).
\]
Under these assumptions, together with the usual positivity conditions, 
the causal target parameter is identified by
\[
\Psi(t_0) = 
E\Big[E\Big[\cdots E\Big[E\bigl[Y(t_0) \mid \bar{L}(t_0), 
\bar{C}(t_0)=1, \Delta=1\bigr] \mid \bar{L}(t_0-1), 
\bar{C}(t_0-1)=1\Big] \cdots \mid L(0)\Big]\Big].
\]
Compared to the right-censoring-only identification result, the innermost 
expectation now additionally conditions on $\Delta = 1$. We discuss 
estimation of this quantity via two approaches in 
Section~\ref{subsec:estimation}.

\medskip
\noindent\textit{In our resampling application, let 
$Y = \mathbb{I}(T > t_0,\, \bar{C}(t_0)=1)$ denote the 
right-censored survival outcome at $t_0$, observed only for 
participants with $\Delta=1$. The observed outcome is then 
$Y^* = \Delta Y = \mathbb{I}(T > t_0,\, \bar{C}(t_0)=1,\, \Delta=1)$. 
The quantity of interest is $\Psi(t_0) = E\bigl[Y^{\bar{C}(t_0)=1}\bigr] 
= P(T > t_0)$; since $\Delta$ affects only what we observe rather than 
the underlying outcome, $Y^{\bar{C}(t_0)=1} = 
Y^{*\,\Delta=1,\,\bar{C}(t_0)=1}$, so both formulations are equivalent. 
The randomization assumption on $\Delta$ holds by design, as resampling 
probabilities are set by study administrators (potentially conditional 
on the participant's observed past); in our application, resampling 
occurs after each participant's administrative end of follow-up $\tau$, 
so this assumption specializes to 
$Y^{\Delta=1,\,\bar{C}(t_0)=1} \perp \Delta \mid V$, where 
$V = (W, \tau, \bar{L}(\tau))$ is the stage-one data observed for 
all participants.}

\subsection{Resampling design-specific notation}
\label{subsec:notation}

We now introduce notation specific to our motivating resampling application; 
this notation is used throughout the estimation section. Let $W$ denote baseline covariates collected on each participant. Let $\tau$ represent some right censoring time (potentially due to moving out of the region, leaving the clinic or study, or simply administrative end of study); $\tau$ can vary for different participants. We may generally refer to tau for simplicity as administrative end of study (as we define it in our simulation setup), but our methods extend to any general right censoring event. For each time point $t$, let $L(t)$ denote time-dependent covariates. We define $L(t) = \bigl(V(t), Id(t), V(t) B(t)\bigr)$
where $V(t)$ is an indicator of attending a clinic visit at time $t$, $Id(t)$ is an indicator that death has occurred and has been reported directly to the clinic at time $t$, and $B(t)$ is a vector of $j$ time-dependent covariates $B(t)=(B_1(t),...,B_j(t))$ measured at time $t$ if a visit occurs (i.e., when $V(t) = 1$). If a visit does not occur at a given time point, we do not collect any time-dependent covariate information and $B(t)$ are all zero, hence we write $V(t) B(t)$. 

Let $T$ denote the time of death. Let $R$ be the resampling indicator; in our example relating to estimating HIV-related mortality resampling may occur once per participant at their respective end of study $\tau$. Only participants who are LTFU are eligible for resampling; here, we define LTFU as having an unknown outcome status at end of study $\tau$. $R$ can be set randomly, such that a fixed proportion of LTFU participants are selected for second-stage sampling, or depending on 
observed covariates to boost efficiency \citep{an2015choosing}. After resampling is complete, let $\Delta$ denote an indicator that a participant’s outcome status is known at the end of follow-up $\tau$. In the context of two-stage design framework, $\Delta$ is an indicator that a participant has second stage information measured. Thus, $\Delta = 0$ if the participant’s outcome remains unknown at $\tau$, even after resampling. $\Delta = 1$ if a participant's outcome at $\tau$ is known, which can occur if: (i) the participant is observed to be alive through $\tau$ (i.e., $V(\tau)=1$), (ii) death is reported to the clinic at or before $\tau$ (i.e., $Id(t) = 1$ for some $t \leq \tau$), or (iii) the participant is selected for resampling ($R = 1$) and their outcome status is subsequently ascertained. Importantly, $\Delta=1$ can occur even if a participant is not chosen for resampling because their outcome status at their end of study $\tau$ may already be deterministically known given their covariate information $(W, \tau, \bar{L}(\tau))$. Such participants are not eligible for resampling as we already know their relevant outcome status. This represents a slight deviation from the more typical two-stage design setup wherein second stage (in our case, outcome) information is only ascertained through collection of additional data in stage two; here, stage one covariates can imply the outcome directly without resampling. 

To bridge notation with the two-stage design framework, we can define the full data as
\[
X = \bigl(W, \tau, \bar{L}(\tau), \bar{Y}(\tau)\bigr).
\]
We note that the full data is typically defined as a complete measurement scenario without censoring; we deviate from this convention by defining $\bigl(W, \tau, \bar{L}(\tau), \bar{Y}(\tau)\bigr)$ as a right-censored data structure. This is a deliberate choice of convenience which facilitates the construction of the estimator we present in ~\ref{subsec:ipcw-ltmle}. The stage-one observed data include baseline covariates $W$, participant end of study $\tau$, and time-dependent covariates $\bar{L}(\tau)$ tracked on each participant through their end of study
\[
V = \bigl(W, \tau, \bar{L}(\tau)\bigr).
\]
Among individuals with $\Delta = 1$, we additionally observe the outcome process $\bar{Y}(\tau)$ for each participant, through their end of study $\tau$. 
The observed data structure 
\[
O = \bigl(W, \tau, \bar{L}(\tau), \Delta, \Delta \bar{Y}(\tau)\bigr)
\]
is not a right censored survival data structure, like the full data $X$, as participants may become LTFU before either death or end of study occurs. 

We note this is a unique case of a two-stage design with outcome subsampling as the amount of information collected in stage two $\bar{Y}(\tau)$ directly depends on data collected in stage one $V$ through the variable $\tau$. For example, if $\tau=2$, we observe $\bar{Y}(2)$ two months (or applicable unit) of survival information on a participant; if $\tau=10$, we observe $\bar{Y}(10)$ ten months of survival information on a participant. It may indeed be more natural to write the full data as $X=(W, \bar{L}(t_0), \bar{Y}(t_0))$ with bivariate censoring by $(\tau, \Delta)$ and define our observed data under coarsening at random (CAR) as $\Phi(X,\tau, \Delta)$; we present an estimator for survival resampling desings motivated by this framework elsewhere \citep{landsiedel2025hazard}. As mentioned, these types of bivariate censored data problems do not admit closed form efficient influence curves or estimators \citep{van1996efficient, quale2002locally}. Thus, we return to our original definition of full data $X = \bigl(W, \tau, \bar{L}(\tau), \bar{Y}(\tau)\bigr)$ as it nicely connects our data structure to the two-stage design formulation and resulting estimators.

\subsection{Estimation}
\label{subsec:estimation}

The estimators proposed in this paper are general to causal parameters 
expressible as marginal means in longitudinal two-stage designs with 
outcome subsampling, as characterized in Section~\ref{subsec:identification}. 
For concreteness, we present both estimators in the notation of our 
motivating resampling application; resampling-specific implementation 
choices, such as the handling of deterministic outcomes, are called out explicitly where they 
arise. The core estimation ideas, however, apply directly to the general class. 

We begin by reviewing how causal parameters expressed in the iterated expectation 
representation of the longitudinal g-computation formula can be estimated 
and targeted through sequential regression paired with sequential targeting 
steps \citep{vdlaan2011mtp, vdlaan2018targeted, petersen2014targeted}, then we introduce our two novel estimators for two-stage designs with outcome subsampling: (1) a targeted IPCW-LTMLE 
and (2) an LTMLE that directly incorporates the second-stage sampling 
mechanism as an intervention node, avoiding explicit inverse weighting 
entirely.

\subsubsection{Overview of LTMLE and sequential regression}
\label{subsec:ltmle_over}

Longitudinal targeted maximum likelihood estimation (LTMLE) combines 
sequential regression with targeted updating to estimate causal target 
parameters that can be expressed as marginal means \citep{vdlaan2011mtp, 
vdlaan2018targeted, petersen2014targeted, shirakawa2024longitudinal, bang2005doubly, van2011targeted}. At 
a high level, LTMLE proceeds via backward sequential regression under a specified intervention regime $\bar{d}$. Let $\bar{Q}_t^d(\cdot)$ denote 
the conditional expectation of the outcome (or future pseudo-outcome) at 
step $t$ under intervention $\bar{d}$. We begin at the last time point, 
regressing the final outcome $Y(K)$ on past covariate and treatment 
history among individuals who follow the regime (i.e., 
$\bar{A}(t)=\bar{d}_t$) at all previous time points, yielding the 
innermost regression $\bar{Q}_K^d(\bar{L}(K-1))$. This model is then 
used to generate predicted outcomes for all individuals under the 
intervention $\bar{d}$. These predicted values are treated as 
pseudo-outcomes in the next regression, where we step back one time point 
and regress on a reduced history. This process is iterated backward 
through time until reaching a regression on baseline covariates alone, 
$\bar{Q}_1^d(L(0))$, at which point the target parameter is obtained by 
taking the empirical mean of the resulting targeted predictions.

The LTMLE estimator also involves updating (or targeting) the initial 
regression estimates at each step of the backward recursion so that the 
resulting estimator solves the efficient influence curve equation for the 
parameter of interest. For a target parameter $\Psi_d(P)$, the efficient 
influence curve is given by
\begin{align*}
D_d^*(P)(O)
&= \bar{Q}_1^d(L(0)) - \Psi_d(P) \\
&\quad + \sum_{t=1}^{K}
\frac{\mathbb{I}\bigl(\bar{A}(t-1) = \bar{d}_{t-1}\bigr)}
{\bar{g}_{t-1}\bigl(A(t-1), \bar{L}(t-1)\bigr)}
\times
\left[
\bar{Q}_{t+1}^d\bigl(\bar{L}(t)\bigr)
-
\bar{Q}_t^d\bigl(\bar{L}(t-1)\bigr)
\right],
\end{align*}
where $\bar{Q}_K^d(\bar{L}(K-1)) = E\bigl[Y(K) \mid \bar{L}(K-1), 
\bar{A}(K-1) = \bar{d}_{K-1}\bigr]$ is the innermost regression and 
$\bar{Q}_{K+1}^d \equiv Y(K)$ is the observed terminal outcome. In survival settings, interventions 
are often defined on the censoring mechanism, where $C(t)$ is treated 
analogously to $A(t)$ and survival is estimated under the intervention 
$C(t)=1$ for all $t \leq t_0$ (i.e., no censoring through $t_0$). The 
LTMLE framework allows for a single outcome $Y$ or an outcome $Y(t)$ 
measured repeatedly over time (like an indicator of having survived 
through $t$ or of having died at or before $t$).

\subsubsection{IPCW-LTMLE for longitudinal two-stage designs}
\label{subsec:ipcw-ltmle}

We first extend the IPCW-TMLE estimator for two-stage designs in the point treatment setting \citep{rose2011targeted} to the longitudinal setting. Recall these estimators call for inverse weighting a ``full data" TMLE, where the ``full data" is collected only on participants chosen for second stage data collection. In our application to resampling, the full data correspond to a right-censored, longitudinal survival data structure
\[
X = (W, \tau, \bar{L}(\tau), \bar{Y}(\tau)),
\]
where $Y(t) = \mathbb{I}(T > t)$ is an indicator of having survived past time $t$. At stage one, we observe
\[
V = (W, \tau, \bar{L}(\tau)),
\]
for all participants, while the survival process $\bar{Y}(\tau)$ is only observed for individuals with $\Delta = 1$. The observed data can therefore be written as
\[
O = (V, \Delta, \Delta \bar{Y}(\tau)).
\]
Note that the full data $X$ can be identically expressed in the longitudinal form defined in section~\ref{subsubsec:right_censoring}. Following \citet{rose2011targeted}, we construct an IPCW-TMLE by applying an LTMLE for the full data structure $X$, combined with inverse probability weights to account for outcome missingness induced when $\Delta=0$. Specifically, each observation is weighted by
\[
\frac{\Delta}{\Pi_{\Delta}} = \frac{\Delta}{P(\Delta = 1 \mid W, \tau, \bar{L}(\tau))}.
\]
These weights are equal to one for individuals whose outcome status is known directly from stage-one data, equal to the inverse of the resampling probability when $R=1$, and equal to 0 if the participant has unknown outcome even after second stage data collection.

We implement this approach using LTMLE for right-censored survival data as our full data estimator \citep{vdlaan2011mtp, vdlaan2018targeted, cai2020one, lendle2017ltmle}, leveraging the sequential regression framework estimation procedure above -- we estimate counterfactual survival under an intervention which prevents censoring (by administrative end of study, induced by $\tau$) through time $t_0$. 

Under standard conditions, IPCW-TMLE is semiparametric efficient when the full-data model is nonparametric and the sampling mechanism is estimated flexibly \citep{rose2011targeted}. In our resampling setting, the censoring time $\tau$ is often known by design and may vary across individuals, so the full-data model is not fully nonparametric. As a result, the estimator is not guaranteed to achieve the nonparametric efficiency bound. 

Additionally, we work here under the sequential randomization assumption (SRA), which implies that censoring affects the underlying full data, i.e. the incorrect assumption that censoring can alter the underlying counterfactual values of the outcome like a treatment. The honest choice would be to assume coarsening at random (CAR), which assumes only that censoring only hides the full data without altering it. As a result, even with a nonparametric full data model and nonparametric estimation of $\Pi$, the IPCW-LTMLE is not efficient in bivariate censored data settings (where you have right censoring in addition to second stage sampling): an estimator that is efficient for the larger SRA model will not generally be efficient for the smaller CAR model  \citep{van1996efficient, quale2002locally}. This is precisely the additional source of inefficiency that arises in bivariate censored data settings when censoring processes are treated like treatment nodes under a causal inference framework \cite{laan2003unified}. However, working under the causal inference framework and assuming SRA opens the door to much easier estimation procedures. Additionally, we can attempt to make these causal inference-inspired estimators more efficient through flexible or targeted estimation of the sampling probabilities $\Pi_{\Delta}$, even when these probabilities are known by design, as is the case in typical two-stage design settings.\\

\noindent\textbf{Estimating and targeting known IPCW weights to gain efficiency}

\noindent We consider two approaches to improve the efficiency of our proposed IPCW-LTMLE estimator: (1) flexible estimation of the sampling probabilities using data-adaptive methods such as Super Learner \citep{van2007super}, and (2) additional targeted estimation of the sampling mechanism itself.

The standard IPCW-LTMLE estimator relies on running LTMLE on observations with $\Delta = 1$, using known inverse probability weights to account for the second-stage sampling. Though known inverse weights account for bias, this estimator can be quite inefficient as data from individuals with $\Delta = 0$ would not be used at all. In other words, even when the true sampling probabilities are known by design, they are not necessarily optimal for variance reduction. By estimating or targeting the sampling probabilities $\Pi_{\Delta}(W,\bar{L}(\tau),\tau) = P(\Delta = 1 \mid W, \bar{L}(\tau), \tau)$, we incorporate information from all participants, and may substantially reduce the variance of our estimator.

Estimation of the inverse weights proceeds by fitting a model for $P(\Delta = 1 \mid W, \tau, \bar{L}(\tau))$ and substituting the estimated probabilities into the IPCW weights. In contrast, targeted estimation introduces an additional updating step designed to enforce a mean-zero efficient influence curve.

When the sampling mechanism $\Pi_{\Delta}$ is estimated and targeted, the influence curve of the resulting IPCW-LTMLE estimator can be written as the inverse-weighted full-data influence curve minus its projection onto the tangent space of the sampling mechanism \citep{rose2011targeted}.

Let $D^F$ denote the full-data efficient influence curve defined in Section~\ref{subsec:ltmle_over}. Here, by ``full data" we refer to our definition $X=(W, \tau, \bar{L}(\tau), \bar{Y}(\tau))$, which is right censored, longitudinal survival data. $D^F$ is the influence curve we get from running LTMLE on observations with $\Delta=1$. The influence curve of the targeted IPCW-LTMLE estimator is then
\begin{align*}
D^{\hat{\Pi}_{\Delta}}(O)
&=
\frac{\Delta}{\Pi_{\Delta}} D^F -
\frac{
E\!\left[D^F \mid W, \bar{L}(\tau), \tau, \Delta=1 \right]
}{
\Pi_{\Delta}
}
\bigl(\Delta - \Pi_{\Delta})
\end{align*}
where the second term we subtract off is the projection of the full-data influence curve onto the tangent space of the sampling mechanism.

The goal of targeting is to update the initial estimate $\hat{\Pi}_{\Delta}$ so that the empirical mean of this influence curve is approximately zero:
\[
\frac{1}{n}\sum_{i=1}^{n} D^{\hat{\Pi}_{\Delta}}(O_i) \approx 0.
\]

Next we introduce the concrete steps required to implement the targeted 
IPCW-LTMLE estimator. Throughout the remainder of the paper, we call
certain observations or outcomes \textit{deterministic}. By this, we refer to participants in our HIV-related mortality example whose outcomes are 
fully known by way of their observed covariates $(W, \tau, \bar{L}(\tau))$ and therefore have $\Delta = 1$ prior to second-stage sampling. For example, a 
participant who makes a clinic visit at their end-of-study time $\tau$ is 
known to be alive at all time points $t \leq \tau$. Additionally, a participant 
whose death was reported to the clinic, i.e.\ $Id(t) = 1$ for some 
$t \leq \tau$, has a known time of death $T$, which fully defines their $\bar{Y}(\tau)$. Both types of participants have 
$\Delta = 1$ by construction before second stage data collection occurs. Only participants with unknown outcomes after accounting for covariate information are eligible for second-stage data collection. While this particular discussion of deterministic outcomes is specific to resampling designs, deterministic information on the outcome or other variables is a common occurrence in general longitudinal studies. \\

\noindent\textbf{Algorithm: IPCW-LTMLE with targeted estimation of $\Pi_{\Delta}$}
\noindent The steps for applying the IPCW-LTMLE estimator with targeted estimation of the second-stage sampling mechanism $\Pi_{\Delta}$ are summarized as follows:

\begin{enumerate}
    \item \textbf{Initial estimation of $\Pi_{\Delta}$.}  
    Obtain an initial estimate of the second stage sampling mechanism given observed covariates from stage one $\hat{\Pi}_{\Delta} = P(\Delta = 1 \mid W, \bar{L}(\tau), \tau)$ using participants who have a non-deterministic value of $\Delta$.  
    These are individuals whose $\Delta$ is not deterministically known from their covariate history and are therefore subject to resampling as defined in detail in Section~\ref{subsec:notation}. We recommend using Super Learner or other flexible estimator for this step.

    \item \textbf{Initial estimation of the full-data influence curve.}  
    Run an initial IPCW-LTMLE (LTMLE with inverse weights $\frac{\Delta}{\hat{\Pi}_{\Delta}}$) to obtain an estimate of the inverse weighted full-data influence curve $\frac{\Delta}{\hat{\Pi}_{\Delta}} \, \hat{D}^F$. Multiply by the inverted weights to isolate $\hat{D}^F$.

    \item \textbf{Estimate the projection.}  
    Among non-deterministic participants with $\Delta = 1$, regress $\hat{D}^F$ on $(W, \bar{L}(\tau), \tau)$ to estimate
    \[
    \widehat{E}\!\left[\hat{D}^F \mid W, \bar{L}(\tau), \tau, \Delta=1 \right].
    \]
    Use this fitted model to predict values for all non-deterministic participants.

    \item \textbf{Construct the clever covariate.}  
    Define
    \[
    H(W, \bar{L}(\tau), \tau)
    =
    \frac{
        \widehat{E}\!\left[\hat{D}^F \mid W, \bar{L}(\tau), \tau, \Delta = 1\right]
    }{
        \hat{\Pi}_{\Delta}
    },
    \]
    for all participants.

    \item \textbf{Targeting step for the resampling mechanism.}  
    Regress $\Delta$ on the clever covariate $H(W, \bar{L}(\tau), \tau)$ using a logistic regression with offset $\text{logit}(\hat{\Pi}_{\Delta})$ among participants with non-deterministic $\Delta$. This step estimates a fluctuation parameter $\hat{\epsilon}$ (the coefficient in from of the clever covariate in the logistic regression) such that the updated weights satisfy the empirical mean‐zero condition for influence curve equation. Specifically, we run the logistic regression
\[
\text{logit}\!\big(\hat{\Pi}_{\Delta}^{\epsilon}\big)
=
\text{logit}\!\big(\hat{\Pi}_{\Delta}\big)
+
\epsilon \, H(W, \bar{L}(\tau), \tau),
\]
where $\hat{\epsilon}$ is the unique solution to the score equation
\[
\frac{1}{n} \sum_{i=1}^{n} H(W_i, \bar{L}_i(\tau), \tau_i)
\left(\Delta_i - \hat{\Pi}_{\Delta,i}^{\epsilon}\right) = 0.
\]
which corresponds exactly to the component of the influence curve arising from fluctuating the resampling mechanism $\Pi_\Delta$, ensuring that this term contributes zero mean to the full influence curve equation.

    \item \textbf{Update the initial estimates of $\hat{\Pi}_{\Delta}$.}  
    Update the initial estimate according to
    \[
    \text{logit}\!\big(\hat{\Pi}_{\Delta}^{\,*}\big)
    =
    \text{logit}\!\big(\hat{\Pi}_{\Delta}\big)
    +
    \hat{\epsilon}\, H(W, \bar{L}(\tau), \tau),
    \]
    yielding the targeted estimate $\hat{\Pi}_{\Delta}^{\,*}$. Here, $\hat{\Pi}_{\Delta}$ denotes the initial estimate, while $\hat{\Pi}_{\Delta}^{\,*}$ denotes the targeted estimate.

    \item \textbf{Compute the updated influence curve.}  
    Substitute $\hat{\Pi}_{\Delta}^{\,*}$ into the influence curve to obtain
    \[
    D^{\hat{\Pi}_{\Delta}^{\,*}}(O)
    =
    \frac{\Delta}{\hat{\Pi}_{\Delta}^{\,*}} \, \hat{D}^F
    -
    \frac{
        \widehat{E}\!\left[\hat{D}^F \mid W, \bar{L}(\tau), \tau, \Delta = 1\right]
    }{
        \hat{\Pi}_{\Delta}^{\,*}
    }
    (\Delta - \hat{\Pi}_{\Delta}^{\,*}).
    \]

    \item \textbf{Iterate until convergence.}  
Repeat the targeting step (Steps 5–7) until convergence, each time plugging-in the newly updated $\hat{\Pi}_{\Delta}^{\,*}$, ensuring that the empirical mean of the updated influence curve is approximately zero, i.e.,
\[
\frac{1}{n} \sum_{i=1}^n D^{\hat{\Pi}_{\Delta}^{\,*}}(O_i) \approx 0,
\]
or practically we can use the following convergence criteria,
\[
\big|\text{mean}\{D^{\hat{\Pi}_{\Delta}^{\,*}}\}\big|
\le
\frac{
    \text{sd}\{D^{\hat{\Pi}_{\Delta}^{\,*}}\}
}{
    \sqrt{n}\,\log(n)
}.
\]
\end{enumerate}

Inference can be based on the influence curve using either a conservative or accurate approach (the accurate approach accounts for efficiency gains stemming from targeting of the IPCW weights). In both cases, we evaluate the influence curve using the true sampling probabilities $\Pi_0$ when they are known by design. While estimating or targeting $\Pi_\Delta$ improves the efficiency of the point estimator, the true sampling probabilities provide a more accurate and stable estimate of the influence curve for variance estimation.

A conservative approach to inference uses the inverse-weighted influence curve:
\[
D^{\mathrm{cons}}(O)
=
\frac{\Delta}{\Pi_0} D^F.
\]

An accurate (non-conservative) approach uses the projected influence curve:
\[
D^{\mathrm{acc}}(O)
=
\frac{\Delta}{\Pi_0} D^F
-
\frac{
E[D^F \mid W, \tau ,\bar{L}(\tau), \Delta=1]
}{
\Pi_0
}
(\Delta - \Pi_0).
\]

The asymptotic variance of the estimator can then be estimated by the empirical variance of the estimated influence curve:
\[
\hat{\sigma}^2
=
\frac{1}{n} \sum_{i=1}^n
D(O_i)^2.
\]
where $D(O)$ denotes either $D^{\mathrm{cons}}(O)$ or $D^{\mathrm{acc}}(O)$.

In practice, $D^{\mathrm{cons}}$ yields conservative inference, while $D^{\mathrm{acc}}$ typically provides more accurate (lower variance) inference when the projection term is well estimated. We demonstrate additional cross-fitted procedures for influence-curve based inference in section~\ref{sec:cross}

\subsubsection{LTMLE for two-stage designs with outcome subsampling}
\label{subsec:novel_ltmle}

Rather than handling censoring by $\tau$ within LTMLE and subsequently 
applying inverse probability weighting to address missingness due to 
$\Delta$, we propose an alternative approach that incorporates both 
missingness mechanisms directly into a single sequential regression 
procedure. This novel implementation of LTMLE for two-stage designs with 
outcome subsampling allows us to return to plug-in principles, critically 
avoiding the use of any direct inverse weighting which may further improve 
the efficiency and stability of the estimator. 

We represent the observed data in longitudinal form
\[
O = \bigl(L(0), C(1), L(1), C(2), \ldots, C(K), L(K), \Delta, \Delta Y(t_0)\bigr),
\]
where $C(t)$ denotes censoring indicators (related to $\tau$) and $\Delta$ 
indicates whether the outcome is observed after resampling, as presented 
in section~\ref{subsubsec:right_censoring_ltfu}. Our final outcome can be defined as
\[
Y(t_0) = \mathbb{I}\bigl(T > t_0,\, \bar{C}(t_0)=1\bigr),
\]
where $t_0$ is the time of interest for survival estimation, so that 
the terminal outcome node in the data structure $\Delta Y(t_0) \equiv Y^*$ 
as introduced in Section~\ref{subsubsec:right_censoring_ltfu}. In our 
resampling example, we target counterfactual survival under joint 
interventions which prevent censoring through $t_0$ and ``set" $\Delta=1$ 
or enforce that all participants are selected for second stage measurement. 
Importantly, incorporating $\Delta$ as an additional intervention node 
allows us to avoid inverse weighting, unlike the IPCW-LTMLE estimators 
we presented previously and the wKM estimators commonly used in our 
motivating example. 

Note that this implementation of LTMLE requires all intermediate $Y$ 
nodes to be removed, retaining only the terminal outcome node $Y(t_0)$. As 
established in Section~\ref{subsubsec:right_censoring_ltfu}, $\Delta$ 
precedes $Y(t_0)$ in the time-ordered node sequence, reflecting that outcome 
ascertainment occurs only after second-stage sampling is complete at the 
end of main study period. Retaining intermediate $Y(t)$ nodes defined 
prior to $\Delta$ would misrepresent this causal and temporal structure 
of the data; however, in our resampling example, information on the outcome that can be gleaned via covariates is still included in $L(t)$ nodes where applicable. 

To implement this estimator, we rewrote the existing \texttt{ltmle} 
software \citep{schwab2023ltmle, petersen2014targeted, lendle2017ltmle, bang2005doubly} for several 
reasons. LTMLE typically relies on Super Learner, a cross-validated 
ensemble machine learning algorithm, to fit the sequential regressions 
and other nuisance parameters \citep{van2007super, Polley2023}. 
Typically, this is an ideal choice which circumvents strong parametric 
modeling assumptions while relying on cross-validation to prevent 
overfitting from flexible learners. However, in our motivating example, 
we have large amounts of deterministic information in the data. 
Deterministic observations (those who are known to be alive or dead from 
covariate information alone) cannot be used to fit the sequential outcome 
regressions. This can dramatically reduce the effective sample size at 
different time points. Thus, we wrote custom software which allows for 
time-varying SuperLearner libraries, which even revert back to 
non-cross-validated lasso regressions when data is too sparse at a given 
time point to support cross-validation. Second, we need to carefully 
leverage the deterministic information in the data when fitting the 
sequential regressions. In particular:
\begin{itemize}
    \item Individuals with $V(t_0)=1$ (equivalently $M \ge t_0$) are 
    known to be alive at time $t_0$, and therefore have $Y=1$ 
    deterministically.
    \item Individuals with $Id(t)=1$ for some $t \le t_0$ are known to 
    have died prior to $t_0$, and therefore are deterministically known 
    to be dead at $t \geq T$.
\end{itemize}
These individuals are excluded from regression steps where the outcome 
is already known, but their deterministic values are retained when 
constructing predictions, which actually stabilized the estimator 
overall. In estimating the resampling mechanism, we also incorporate 
deterministic knowledge of $\Delta$. In particular, $\Delta=1$ is 
assigned deterministically for individuals whose outcome status is known 
from observed data as described previously. The LTMLE package does have 
deterministic Q and g arguments, which may be able to appropriately 
handle deterministic information, depending on the setting. Last, we 
needed custom sequential regression code in order to implement the 
cross-fitted variance estimator presented in section~\ref{sec:cross}. 

From there, implementing this LTMLE follows the sequential regression 
procedure defined in section~\ref{subsec:ltmle_over}. At the final time 
point, we regress $Y$ on $\bar{L}(t_0)$ among individuals who follow the intervention 
regime (i.e., $\bar{C}(t_0)=1$ and $\Delta=1$) and whose outcomes are 
not deterministically known.
Predictions from this model are generated for all individuals 
under the aforementioned interventions, with deterministic 
outcomes set to 1 (known alive) or 0 (known dead). These 
predicted values serve as pseudo-outcomes in the next step 
of the backward recursion. At each step $t = K, 
K-1, \ldots, 1$, we regress the current 
pseudo-outcome on past history $\bar{L}(t-1)$ among individuals who 
follow the intervention regime up to time $t$ and whose outcome is not 
deterministically known, incorporating deterministic information before 
each targeting step. Predictions are then generated for all individuals, 
targeted, and propagated backward as pseudo-outcomes, yielding a sequence 
of iterated conditional expectations. The final estimator is obtained by 
taking the empirical mean of the predictions from the regression on 
baseline covariates $L(0)$, as in standard LTMLE.

Although we have presented both estimators in the notation of 
the resampling setting, the core ideas extend directly to any 
two-stage design with outcome subsampling as defined in 
Section~\ref{subsec:identification}. In the general setting, 
the IPCW-LTMLE applies a full-data LTMLE among participants 
with $\Delta=1$, inversely weighted by $P(\Delta=1 \mid V)$, 
where $V$ denotes whatever stage-one data is available. The 
full LTMLE incorporates $\Delta$ as an additional intervention 
node in the sequential regression, with $Y(t_0)$ defined as the 
terminal outcome downstream of $\Delta$. Resampling-specific 
implementation choices such as deterministic outcome handling 
do not affect the core estimation logic and can be omitted in 
settings where they do not arise.
\subsection{Cross-fitted variance estimation}
\label{sec:cross}

Initial simulation studies revealed substantial undercoverage of influence-curve-based confidence intervals for both the IPCW-LTMLE and the full LTMLE in certain settings. We attribute this to overfitting of the sequential outcome regressions, which are estimated on the same data used to evaluate the influence curve. For the IPCW-LTMLE, the projection term $E[D^F \mid W, \bar{L}(\tau), \tau]$ is an additional 
source of overfitting. When flexible, data-adaptive methods such as Super Learner are used to estimate these nuisance components, the resulting influence curve evaluations are systematically optimistic, leading to underestimation of variability and anticonservative inference \citep{zheng2010asymptotic, j2025performance, zivich2021machine, chernozhukov2018double, an2009need}. To address this, we propose cross-fitted variance estimators for both the IPCW-LTMLE 
and the full LTMLE. Our approach retains the original point estimator, computed on the full data, and applies sample splitting only to the variance estimation step, 
re-estimating the sequential regressions and, for the IPCW-LTMLE, the projection term, on held-out folds.

This design choice is deliberate. The resampling data structure considered here contains substantial deterministic information, which reduces the effective sample size available for nuisance estimation, particularly at smaller sample sizes and early time points. A fully cross-fitted procedure, in which the point estimate itself is also computed via sample splitting, would further reduce the effective sample size for point estimation and risk destabilizing the estimator. Since the point estimates were well-behaved in our initial studies, we opted for a hybrid approach: non-cross-fitted point estimation paired with cross-fitted variance estimation. This preserves the stability and efficiency of the point estimator while providing variance estimates that are robust to overfitting. We describe the procedure for each estimator in detail below.

\subsubsection{Cross-fitted variance estimation for IPCW-LTMLE}
\label{subsec:cf_ipcw}

To mitigate the effects of overfitting on confidence interval coverage, we implement cross-fitted estimation of both the conservative and accurate influence curves. Cross-fitting is particularly important in this setting, as both the full-data influence curve component $D^F$ and the projection term $E[D^F \mid W, \tau, \bar{L}(\tau)]$ may be estimated using flexible, data-adaptive methods trained on the same data.\\

\noindent\textbf{Cross-fitted construction of $D^F$}

\noindent We first construct cross-fitted estimates of components of full-data influence curve component $D^F$. We partition the data into $V$ folds. For each fold $v$, we fit the sequential LTMLE outcome regressions $\{\bar{Q}_t\}$ on the training data (excluding fold $v$), using only observations with $\Delta = 1$. The full sequence of regressions (and predictions) are computed entirely within the training data. Note we do not need to target the sequential regressions if we only care about estimating the influence curve. Targeting is meant to provide an optimal bias-variance trade-off for the target parameter of interest; here we only care about a well-estimated influence curve, so targeting is an unnecessary additional step that we forgo. 

Once all sequential regressions have been fit on the training data, we generate predictions on the validation fold under an intervention that eliminates censoring. These predictions are used to construct cross-fitted estimates $D_i^{F,\mathrm{cf}}$ (just the $Q$ terms) for each observation $i$ in fold $v$. Repeating this procedure across all folds yields a cross-fitted estimate $D_i^{F,\mathrm{cf}}$ for every observation, where each value is computed using models fit on data excluding that observation. Note that we did not cross-fit the estimates of the censoring mechanism as our goal was getting a better variance estimator. To that end, we focus on cross-fitting terms which appear as residuals in our influence curve representation. Additional cross-fitting of the censoring/intervention mechanims may be of interest is one wishes to avoid Donsker class conditions \citep{j2025performance}.\\

\noindent\textbf{Cross-fitted projection}

\noindent For the accurate influence curve, we also construct cross-fitted estimates of the projection term
\[
E[D^F \mid W, \tau, \bar{L}(\tau)].
\]
Within each fold, we regress $D^F$ (as computed on the training data only) on $(W, \tau, \bar{L}(\tau))$. We then use the fitted model to generate predictions on the validation fold. This yields cross-fitted estimates $\widehat{E}[D^F \mid W, \tau, \bar{L}(\tau)]^{\mathrm{cf}}$ for all observations. \\

\noindent\textbf{Variance estimators}

\noindent We consider four variance estimators, corresponding to whether cross-fitting is used and whether the projection correction is applied:

\begin{enumerate}

\item \textbf{Conservative (non-cross-fitted):}
\[
\mathrm{IC}_i^{\mathrm{cons}} =
\frac{\Delta_i}{\Pi_0} \, D_i^F,
\]
where $D^F$ is computed using all data.

\item \textbf{Conservative (cross-fitted):}
\[
\mathrm{IC}_i^{\mathrm{cons,cf}} =
\frac{\Delta_i}{\Pi_0} \, D_i^{F,\mathrm{cf}},
\]
where $D^{F,\mathrm{cf}}$ is the cross-fitted version.

\item \textbf{Accurate (non-cross-fitted):}
\[
\mathrm{IC}_i^{\mathrm{accurate}} =
\frac{\Delta_i}{\Pi_0} \, D_i^F
-
\frac{
\widehat{E}[D^F \mid W_i, \tau_i, \bar{L}_i(\tau_i)]
}{
\Pi_0
}
\bigl(\Delta_i - \Pi_0\bigr),
\]
where both $D^F$ and the projection are estimated using all data.

\item \textbf{Accurate (cross-fitted):}
\[
\mathrm{IC}_i^{\mathrm{accurate,cf}} =
\frac{\Delta_i}{\Pi_0} \, D_i^{F,\mathrm{cf}}
-
\frac{
\widehat{E}[D^F \mid W_i, \tau_i, \bar{L}_i(\tau_i)]^{\mathrm{cf}}
}{
\Pi_0
}
\bigl(\Delta_i - \Pi_0\bigr),
\]
where both components are cross-fitted.

\end{enumerate}

The variance of $\hat{\Psi}$ is estimated using the empirical variance of the chosen influence curve:
\[
\widehat{\mathrm{Var}}(\hat{\Psi})
=
\frac{1}{n} \sum_{i=1}^n \mathrm{IC}_i^2.
\]
Note, as mentioned previously, we plug true weights $\Pi_0$ into the IC representation as opposed to estimated or targeted weights. \\

\noindent\textbf{Early time points}

\noindent At early time points where no intervention nodes are present, ie there has not been right censoring yet, the estimator reduces to a weighted mean and does not involve outcome regression. In these cases, the $D^F$ component is not subject to overfitting. However, the projection term $E[D^F \mid W, \tau, \bar{L}(\tau)]$ may still be estimated using flexible methods and therefore remains susceptible to overfitting. Accordingly, we apply cross-fitting only to the projection component in these settings.

Additionally, because we combine cross-fitting with Super Learner for estimation of the sequential regressions, cross-validation is effectively nested within the cross-fitting procedure. This can result in limited sample sizes within training folds, particularly after restricting to non-deterministic observations. To address potential instability due to sparsity, we implement a fallback to a generalized linear model (GLM) when the Super Learner fails to converge or produces errors.

\subsubsection{LTMLE cross-fitted variance estimation}

We partition the $n$ observations into $V$ folds. For each fold $v \in \{1, \ldots, V\}$, we fit the sequential outcome regressions $\{\hat{\bar{Q}}_t^{(-v)}\}$ on the training data (all observations not in fold $v$), proceeding backward in time from $t = K$ to $t = 1$, as in the standard LTMLE procedure but without a targeting step. The censoring mechanism $\hat{g}$ is estimated once on the full data and reused across folds. The procedure for cross-fitting the sequential regression is very similar as presented for the IPCW-LTMLE.

For each observation $i$ in the validation fold $v$, we compute the cross-fitted influence curve contribution by evaluating the influence curve using the nuisance estimates $\{\hat{\bar{Q}}_t^{(-v)}\}$ obtained from the training data. This yields cross-fitted contributions $\hat{D}_i^{\mathrm{CF}}$, where each observation is evaluated using nuisance functions estimated on data excluding that observation.

The cross-fitted variance estimator is then given by
\[
\widehat{\mathrm{Var}}^{\mathrm{CF}}(\hat{\Psi})
=
\frac{1}{n} \sum_{i=1}^n \left(\hat{D}_i^{\mathrm{CF}}\right)^2.
\]

Importantly, the point estimate $\hat{\Psi}$ is always computed using the LTMLE fit on the complete data; only the variance estimation step is cross-fitted.

In some cases, cross-fitting may fail at a specific time point due to data sparsity within a fold (e.g., insufficient observations to fit the sequential regressions using cross-validated SuperLearner). In these cases, which often arise due to the amount of deterministic information contained in this data structure, we fall back to the standard (non-cross-fitted) variance estimator. When this occurs, it typically happens at earlier time points in our simulation studies.

\section{Simulations}
\label{sec:sims}

In this section we compare the performance of the estimators proposed in this paper relative to other estimators proposed in previous literature. This section is organized as follows. First, in section~\ref{subsec:data_gen} we describe the data generating process used for these simulations studies. Second, in section section~\ref{subsec:estimator_comp}, we describe each estimator implemented and details on how inference is computed. Last, in section~\ref{subsec:sim_results} we provide results for the bias, variance, mean squared error (MSE), and 95\% confidence interval coverage for each approach.

\subsection{Data generating processes}\label{subsec:data_gen}
We simulated longitudinal data over $t = 10$ time points across sample sizes $N \in \{500, 1000, 3000\}$. This data includes: baseline covariates, time-varying visit indicators, time-varying CD4 biomarker values (collected only if a visit is made at that $t$), time-varying death indicators (potentially unobserved), time-varying natural death reporting indicators (if a patient dies and has death reported to the clinic on their behalf), and a resampling indicator which triggers the uncovering of (otherwise unobserved) death information if a participant is resampled. Baseline covariates included three binary variables ($W_1, W_2, W_3$) representing risk and protective factors. At each time point, individuals faced a probability of death dependent on their baseline covariates, prior visit behavior, and prior CD4 count values. If death occurred (potentially unobserved) time of death $T$ was generated. If a death occurred, it was naturally reported to the clinic $\text{Id}(t)$ with probability 0.2 among newly deceased individuals at that time point. Visit $V(t)$ probabilites were drawn based on covariates, past visit history, and lagged CD4 counts, with biomarker values $L(t)$ representing CD4 levels evolving autoregressively with Gaussian noise and penalized means based on covariates and visit behavior. Observed CD4 counts were carried forward from previously observed values if a visit was not made at a given time. Each individual was assigned a censoring time $\tau \in \{5,7,9,10\}$ with different probability for each $\tau$, and all data past $\tau$ were censored. The outcome was an indicator of survival by time $t_0$, derived either from observed clinic deaths, observed clinical visits (indicating survival), or resampling. Individuals with unobserved outcomes were chosen for resampling (and subsequent outcome measurement) with probability 0.2. Further information and data generating simulation code can be found in the Supplementary material in section~\ref{sec:supp}.

\subsection{Estimators}\label{subsec:estimator_comp}

\begin{enumerate}
    \item IPCW-LTMLE (known weights $\Pi_{\Delta}$)
    \item IPCW-LTMLE (estimated weights $\hat{\Pi}_{\Delta}$)
    \item IPCW-LTMLE (targeted weights $\hat{\Pi}^*_{\Delta}$)
    \item LTMLE
    \item Hazard TMLE
    \item Weighted Kaplan-Meier (estimated weights $\hat{\Pi}_{\Delta}$)
    \item Weighted Kaplan-Meier (known weights $\Pi_{\Delta}$)
\end{enumerate}

The IPCW-LTMLE estimator (1) applies LTMLE to the right-censored data structure 
$(W, \tau, \bar{L}(\tau), \bar{Y}(\tau))$ among participants with $\Delta = 1$, and accounts for informative outcome missingness using known inverse probability weights $\frac{\Delta}{\Pi_{\Delta}}$. Super Learner was used for the censoring and outcome models, and the library included linear regression, lasso regression, and multivariate regression splines with and without covariate screening \citep{Polley2023}. The inverse weighted full data influence curve is used for variance estimation. 

The IPCW-LTMLE estimator (2) is identical to estimator (1), except that the sampling probabilities $\Pi_{\Delta}$ are estimated among participants with non-deterministic $\Delta$ using Super Learner. The same Super Learner library as for (1) was used for all regression fits, now including the estimation of the sampling probabilities. The inverse weighted full data influence curve is used for variance estimation.

The IPCW-LTMLE estimator (3) further targets the sampling mechanism. As in estimator (2), $\Pi_{\Delta}$ is first estimated using Super Learner. In addition, the projection term $E[D^F \mid W, \bar{L}(\tau), \tau]$ is estimated using Super Learner with the same library. Multiple variance estimates are provided for this estimator as described in the cross-fitting section~\ref{sec:cross}. In the tables, we report the cross-fitted accurate variance, which accounts for the increased efficiency stemming from targeted weight estimation. 

The LTMLE estimator (4), detailed in section~\ref{subsec:novel_ltmle}, incorporates both administrative censoring and outcome missingness directly into the sequential regression procedure. Cross-fit influence curve variance is reported for this estimator.

The hazard TMLE estimator (5) was first proposed in \cite{landsiedel2025hazard}. This TMLE fits a pooled hazard regression among participants with $\Delta=1$ and subsequently targets the conditional hazard for all participants. The product integral representation of survival is inverse weighted to account for censoring by $\tau$. Cross-fitted influence-curve-based variance is used for this 
estimator. We include it as a comparator because it represents 
an alternative TMLE-based approach that avoids direct inverse 
weighting by $\Delta$ but does rely on inverse weighting to 
handle administrative censoring by $\tau$, providing a useful 
intermediate comparison between the full plug-in LTMLE and the 
IPCW-LTMLE estimators.

Estimator (6) is the weighted Kaplan–Meier estimator using estimated resampling probabilities $\hat{\Pi}_{\Delta}$. This estimator assigns weight 1 to individuals with known outcome based on clinical data, weight 0 to individuals with unknown outcome even after resampling, and weight equal to the inverse of the estimated resampling probability to individuals whose outcome was recovered through resampling. Nonparametric bootstrapping is used for inference for this estimator. 

Estimator (7) is the weighted Kaplan–Meier estimator using known resampling probabilities $\Pi_{\Delta}$. This estimator assigns weights as in estimator (6) but relies on the known sampling probabilities. We implement both Kaplan–Meier estimators using the \texttt{survival} package in \texttt{R} \citep{survival-package}. Nonparametric bootstrapping is used for inference for this estimator. 

Point estimates for IPCW-LTMLE were computed using the \texttt{ltmle} \texttt{R} 
package \citep{schwab2023ltmle}. The non-inverse weighted LTMLE, Hazard TMLE, and all 
cross-fitted variance estimators required custom software developed as part of 
this work. Throughout all analyses involving ensemble machine learning, nuisance 
parameters were estimated using the \texttt{SuperLearner} \texttt{R} package 
\citep{polley2025superlearner}.

\subsection{Simulation results}
\label{subsec:sim_results}

In this section, we begin by presenting results for the IPCW-LTMLE estimator when known, versus estimated, versus targeted inverse weights are used. Then we provide an overall estimator comparison, focusing on the targeted version of the IPCW-LTMLE estimator. Last, we demonstrate the coverage performance of our cross-fitted variance estimators for our TMLE-based estimators relative to oracle and non-crosffit performance. Here, oracle coverage refers to confidence interval coverage computed 
using the empirical variance of the point estimate across simulations as 
the true variance; this provides a benchmark for the best achievable 
coverage given the point estimator itself, isolating whether coverage 
failures stem from variance estimation rather than bias in the point 
estimates.

We evaluated five estimators for survival---targeted IPCW-LTMLE, 
LTMLE, Hazard TMLE, and inverse probability weighted Kaplan-Meier (wKM) with 
both known and estimated censoring weights---across sample sizes $N \in \{500, 
1000, 3000\}$. Performance was assessed 
via empirical bias, variance, mean squared error, and 95\% confidence interval 
coverage across ten timepoints over 1,000 Monte Carlo replications. All 
estimators exhibited negligible empirical bias across all sample sizes and 
timepoints, as expected theoretically.

At $N = 500$, LTMLE achieved substantially lower variance than any other 
estimator, particularly at early timepoints: empirical variance of 
$5.5 \times 10^{-5}$ at $t = 1$ compared to $2.0 \times 10^{-4}$ for 
wKM with known weights and $1.5 \times 10^{-4}$ for Hazard TMLE, representing 
reductions of 73\% and 63\% respectively. At $N = 1{,}000$, LTMLE achieved slightly lower variance than Hazard TMLE 
at some timepoints ($5.0 \times 10^{-5}$ vs $6.8 \times 10^{-5}$ at $t = 1$), 
though both remained substantially more efficient than wKM with known weights 
($9.9 \times 10^{-5}$). By $N = 3{,}000$, variance differences among the 
three TMLE methods narrowed considerably, with Hazard TMLE and LTMLE achieving 
comparable efficiency ($2.2 \times 10^{-5}$ and $2.5 \times 10^{-5}$ at 
$t = 1$ respectively), while wKM with known weights remained substantially 
less efficient ($3.2 \times 10^{-5}$).

All three TMLE methods consistently outperformed wKM with known weights across 
all sample sizes and timepoints, with LTMLE achieving variance reductions of 
30--73\% depending on sample size and timepoint. This provides a clear 
efficiency argument over the approach most commonly employed in practice. 
Weighted KM with estimated weights was substantially more competitive than its 
known-weight counterpart. At $N = 1{,}000$ for example, wKM with estimated weights achieved variance of 
$9.8 \times 10^{-5}$ at $t = 1$, nearly identical to wKM with known weights 
($9.9 \times 10^{-5}$), yet by $t = 10$ the gap widened substantially 
($5.3 \times 10^{-4}$ vs $7.1 \times 10^{-4}$). At larger sample sizes, wKM with estimated weights 
approached the efficiency of Hazard TMLE and IPCW-LTMLE (even surpassing it in the tail), though LTMLE 
maintained a variance advantage particularly at earlier timepoints. In particular, at time points ($t = 1$--$4$), the TMLE estimators, in particular the LTMLE and Hazard TMLE, achieve substatially lower variance than the 
weighted Kaplan-Meier estimator with estimated weights, with reductions of 
roughly 15--30\% common at $N \in \{1000, 3000\}$. 

Across all three TMLE methods, the standard non-cross-fitted variance estimator 
substantially underestimated sampling variability, yielding coverage as low as 
80.8\% for Hazard TMLE at $N = 500$, $t = 1$, 86.8\% for LTMLE at $N = 500$, 
$t = 10$, and 76.2\% for IPCW-LTMLE at $N = 500$, $t = 1$. Cross-fitted 
variance estimation restored coverage to near-nominal levels in nearly all 
settings examined.

For IPCW-LTMLE, four variance estimators were compared, reflecting the 
conservative versus non-conservative and cross-fitted versus standard 
formulations. The non-conservative and non-cross-fitted variance estimator performed worst, 
consistently undercovering across timepoints. The non-cross-fitted conservative
estimator recovered somewhat but remained below nominal. Cross-fitted variants 
of both formulations yielded substantially improved coverage: the conservative 
cross-fitted estimator slightly over-covered (97--99\%), while the 
non-conservative cross-fitted estimator most closely tracked the nominal 95\% 
level across sample sizes.

All three estimators exhibit coverage issues at $t = 1$ when $n = 500$, even with 
cross-fitting; at $n = 1{,}000$ and $n = 3{,}000$, coverage in this region is 
near-nominal for all estimators. At $n = 500$, IPCW-LTMLE and hazard-TMLE show 
undercoverage, while LTMLE shows overcoverage. Oracle coverage diagnostics reveal 
distinct failure mechanisms for each estimator. For IPCW-LTMLE and hazard-TMLE, 
oracle coverage is near-nominal, indicating that undercoverage arises from the 
variance estimation rather than from bias in the point estimates 
themselves. The LTMLE presents a contrasting picture: oracle coverage is below 
nominal at $t = 1$, suggesting that the point estimates themselves are affected in 
this region; however, the cross-fitted variance estimator is sufficiently conservative 
to overcompensate, yielding net overcoverage in practice. The low effective sample 
size at $t = 1$, driven by a high degree of deterministic covariate histories 
combined with very few observed events, likely makes both point estimation and 
variance estimation challenging in this region at small sample sizes.

\subsubsection{IPCW-LTMLE}

We begin by presenting results for the IPCW-LTMLE estimator when known weights versus estimated weights versus targeted weights are used for $\Pi_{\Delta}$.

\begin{figure}[H]
  \centering
  \includegraphics[width=\textwidth]{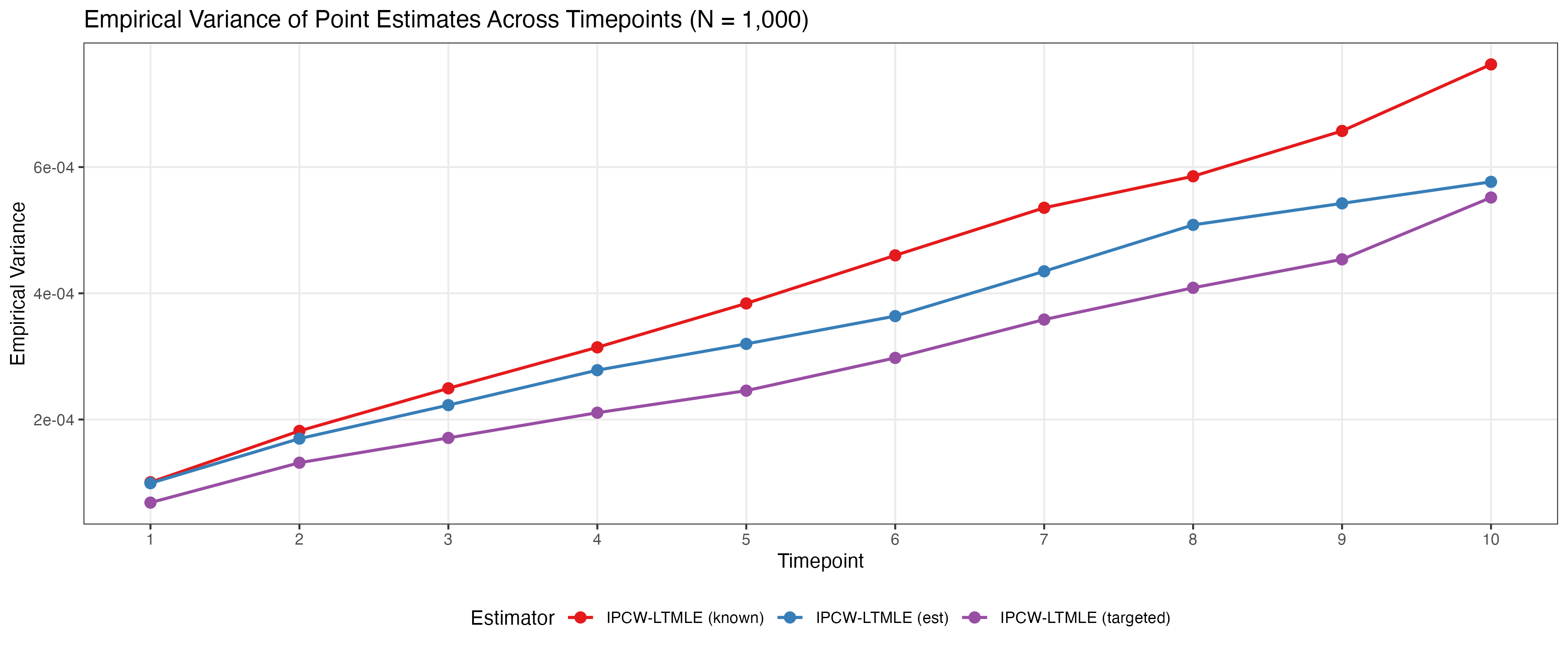}  
\caption{Variance of the point estimate for IPCW-LTMLE with known weights, estimated weights, and targeted weights across 1,000 simulations of sample size N=1,000.}
  \label{fig:ipcw_graph}
\end{figure}

\begin{sidewaystable}
\centering
\caption{Results from 1k simulation studies with sample size $N = 1$k per simulation. Variance reported is the variance of the point estimate across simulations. Coverage for the IPCW-TMLE estimators with known and estimated weights comes from the inverse weighted full data influence curve. Inference for the targeted IPCW-LTMLE estimator is the cross-fitted accurate variance estimator.} 
\label{tab:ipcw_variants_N1000}
\begin{tabular}{llllllllllll}
  \toprule
Metric & Estimator & $t=1$ & $t=2$ & $t=3$ & $t=4$ & $t=5$ & $t=6$ & $t=7$ & $t=8$ & $t=9$ & $t=10$ \\ 
  \midrule
bias & IPCW-LTMLE (known) & 0.000949 & 0.000487 & 0.000303 & 0.000300 & 0.000464 & 0.000492 & 0.000567 & 0.001206 & 0.001523 & 0.001927 \\ 
   & IPCW-LTMLE (est) & 0.000114 & 0.000235 & 0.000193 & 0.000217 & 0.000305 & 0.000366 & 0.000537 & 0.000877 & 0.001886 & 0.002057 \\ 
   & IPCW-LTMLE (targeted) & -0.000861 & -0.000315 & -0.000483 & -0.000573 & -0.000462 & -0.000581 & -0.000538 & -0.000507 & -0.000022 & 0.000706 \\ 
   \midrule
var & IPCW-LTMLE (known) & 0.000101 & 0.000182 & 0.000249 & 0.000314 & 0.000384 & 0.000460 & 0.000536 & 0.000585 & 0.000657 & 0.000763 \\ 
   & IPCW-LTMLE (est) & 0.000099 & 0.000170 & 0.000223 & 0.000278 & 0.000320 & 0.000364 & 0.000435 & 0.000508 & 0.000543 & 0.000577 \\ 
   & IPCW-LTMLE (targeted) & 0.000068 & 0.000132 & 0.000171 & 0.000211 & 0.000246 & 0.000298 & 0.000358 & 0.000409 & 0.000454 & 0.000552 \\ 
   \midrule
mse & IPCW-LTMLE (known) & 0.000101 & 0.000182 & 0.000249 & 0.000314 & 0.000384 & 0.000460 & 0.000535 & 0.000586 & 0.000659 & 0.000766 \\ 
   & IPCW-LTMLE (est) & 0.000099 & 0.000170 & 0.000223 & 0.000278 & 0.000320 & 0.000364 & 0.000435 & 0.000509 & 0.000546 & 0.000580 \\ 
   & IPCW-LTMLE (targeted) & 0.000069 & 0.000131 & 0.000171 & 0.000211 & 0.000246 & 0.000298 & 0.000358 & 0.000409 & 0.000453 & 0.000552 \\ 
   \midrule
coverage & IPCW-LTMLE (known) & 0.895 & 0.925 & 0.943 & 0.948 & 0.937 & 0.937 & 0.926 & 0.938 & 0.939 & 0.928 \\ 
   & IPCW-LTMLE (est) & 0.895 & 0.937 & 0.954 & 0.954 & 0.951 & 0.959 & 0.963 & 0.960 & 0.963 & 0.961 \\ 
   & IPCW-LTMLE (targeted) & 0.941 & 0.925 & 0.936 & 0.935 & 0.933 & 0.935 & 0.940 & 0.934 & 0.941 & 0.950 \\ 
   \bottomrule
\end{tabular}
\end{sidewaystable}

\newpage

\subsubsection{Overall estimator comparison}

Next, we compare the empirical variance, bias, MSE, and coverage of each proposed estimator. Note that from here on, we use the targeted IPCW-LTMLE estimator and report 95\% confidence interval coverage for all TMLE estimators using cross-fitted variance. We show these results across three sample sizes $N \in \{500, 1000, 3000\}$ to highlight the small versus large sample size performance differences among these estimators. 

\subsubsection{Sample size N=500}
\begin{figure}[H]
  \centering
  \includegraphics[width=\textwidth]{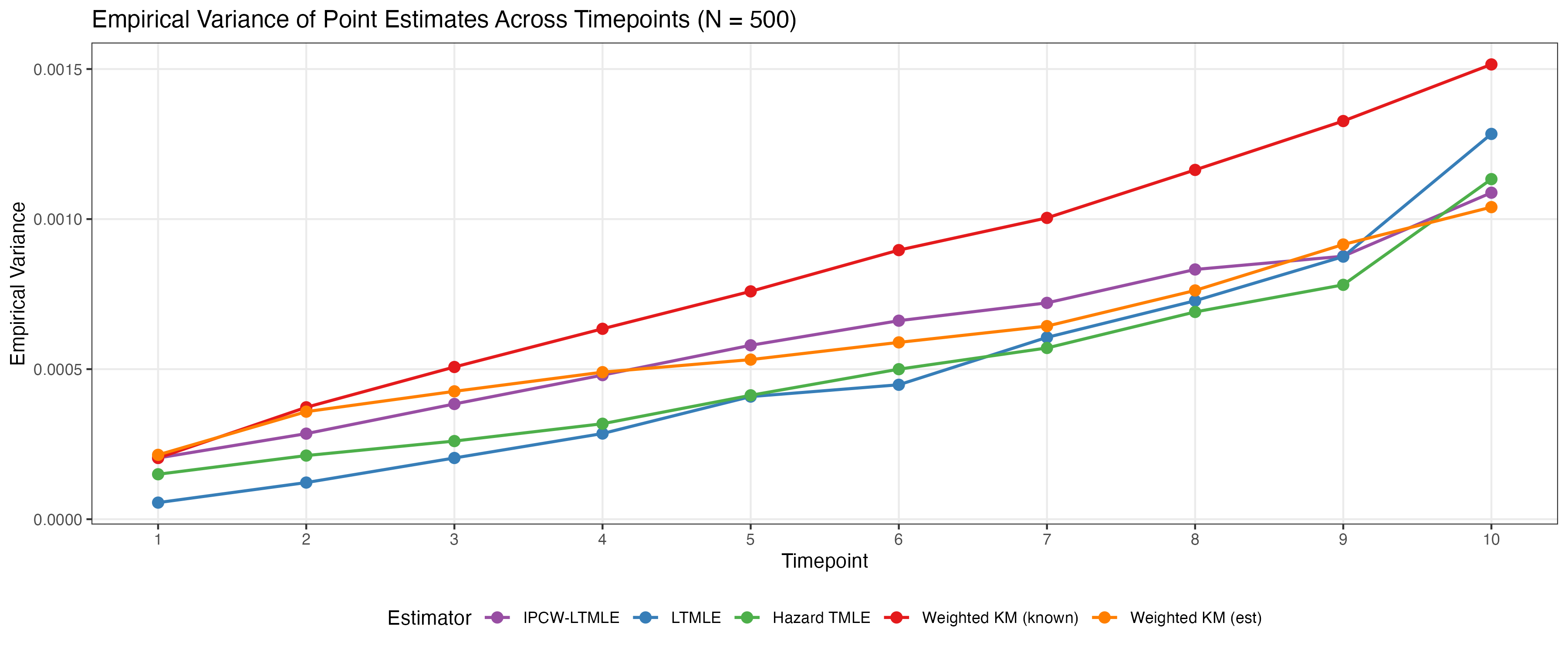}  
\caption{Variance of the point estimate for each estimator across 1,000 simulations of sample size N=500.}
  \label{fig:N500_var_graph}
\end{figure}

\begin{sidewaystable}
\centering
\caption{Results from 1k simulation studies with sample size $N = 500$ per simulation. Variance reported is the variance of the point estimate across simulations. Inference for each estimator which supports 95\% confidence interval coverage is described in Section~\ref{subsec:estimator_comp}. Here, (est) refers to estimated $\Pi_{\Delta}$ and (known) refers to the use of known values of $\Pi_{\Delta}$.} 
\label{tab:results_N500}
\begin{tabular}{llllllllllll}
  \toprule
Metric & Estimator & $t=1$ & $t=2$ & $t=3$ & $t=4$ & $t=5$ & $t=6$ & $t=7$ & $t=8$ & $t=9$ & $t=10$ \\ 
  \midrule
bias & IPCW-LTMLE & -0.001275 & -0.000577 & -0.000445 & -0.000452 & -0.000357 & -0.000476 & -0.000760 & -0.000637 & -0.000593 & 0.001340 \\ 
   & LTMLE & -0.005327 & 0.000800 & 0.003310 & 0.003999 & 0.001276 & 0.004865 & 0.000350 & 0.003904 & -0.001406 & -0.001472 \\ 
   & Hazard TMLE & 0.000119 & 0.000614 & 0.000610 & 0.000265 & 0.000485 & 0.001177 & 0.000437 & 0.000643 & -0.000843 & -0.000700 \\ 
   & Weighted KM (known) & 0.000455 & 0.000549 & 0.000401 & 0.000177 & 0.000165 & -0.000368 & -0.000222 & -0.000043 & 0.000089 & 0.000611 \\ 
   & Weighted KM (est) & 0.000280 & 0.000074 & 0.000420 & 0.000971 & 0.000885 & 0.001096 & 0.000213 & -0.000122 & -0.000240 & -0.000553 \\ 
   \midrule
var & IPCW-LTMLE & 0.000204 & 0.000285 & 0.000384 & 0.000480 & 0.000579 & 0.000661 & 0.000721 & 0.000832 & 0.000876 & 0.001088 \\ 
   & LTMLE & 0.000055 & 0.000122 & 0.000204 & 0.000285 & 0.000409 & 0.000448 & 0.000606 & 0.000728 & 0.000875 & 0.001284 \\ 
   & Hazard TMLE & 0.000150 & 0.000212 & 0.000260 & 0.000318 & 0.000413 & 0.000500 & 0.000571 & 0.000691 & 0.000781 & 0.001133 \\ 
   & Weighted KM (known) & 0.000204 & 0.000373 & 0.000507 & 0.000634 & 0.000759 & 0.000897 & 0.001004 & 0.001164 & 0.001327 & 0.001515 \\ 
   & Weighted KM (est) & 0.000214 & 0.000358 & 0.000426 & 0.000490 & 0.000532 & 0.000589 & 0.000643 & 0.000762 & 0.000915 & 0.001040 \\ 
   \midrule
mse & IPCW-LTMLE & 0.000205 & 0.000285 & 0.000384 & 0.000480 & 0.000579 & 0.000661 & 0.000721 & 0.000832 & 0.000876 & 0.001089 \\ 
   & LTMLE & 0.000083 & 0.000122 & 0.000215 & 0.000301 & 0.000410 & 0.000471 & 0.000605 & 0.000742 & 0.000876 & 0.001285 \\ 
   & Hazard TMLE & 0.000150 & 0.000212 & 0.000260 & 0.000318 & 0.000413 & 0.000501 & 0.000570 & 0.000690 & 0.000781 & 0.001132 \\ 
   & Weighted KM (known) & 0.000204 & 0.000373 & 0.000507 & 0.000634 & 0.000758 & 0.000896 & 0.001003 & 0.001163 & 0.001325 & 0.001514 \\ 
   & Weighted KM (est) & 0.000214 & 0.000358 & 0.000426 & 0.000490 & 0.000532 & 0.000590 & 0.000643 & 0.000761 & 0.000914 & 0.001039 \\ 
   \midrule
coverage & IPCW-LTMLE & 0.779 & 0.918 & 0.921 & 0.929 & 0.912 & 0.942 & 0.947 & 0.948 & 0.951 & 0.939 \\ 
   & LTMLE & 0.984 & 0.970 & 0.934 & 0.922 & 0.937 & 0.930 & 0.933 & 0.916 & 0.939 & 0.928 \\ 
   & Hazard TMLE & 0.860 & 0.936 & 0.950 & 0.951 & 0.943 & 0.932 & 0.948 & 0.952 & 0.965 & 0.969 \\ 
   & Weighted KM (known) & 0.852 & 0.901 & 0.915 & 0.923 & 0.924 & 0.931 & 0.930 & 0.937 & 0.948 & 0.932 \\ 
   & Weighted KM (est) & 0.843 & 0.915 & 0.920 & 0.924 & 0.940 & 0.953 & 0.961 & 0.956 & 0.940 & 0.952 \\ 
   \bottomrule
\end{tabular}
\end{sidewaystable}

\newpage 

\subsubsection{Sample size N=1,000}
\begin{figure}[H]
  \centering
  \includegraphics[width=\textwidth]{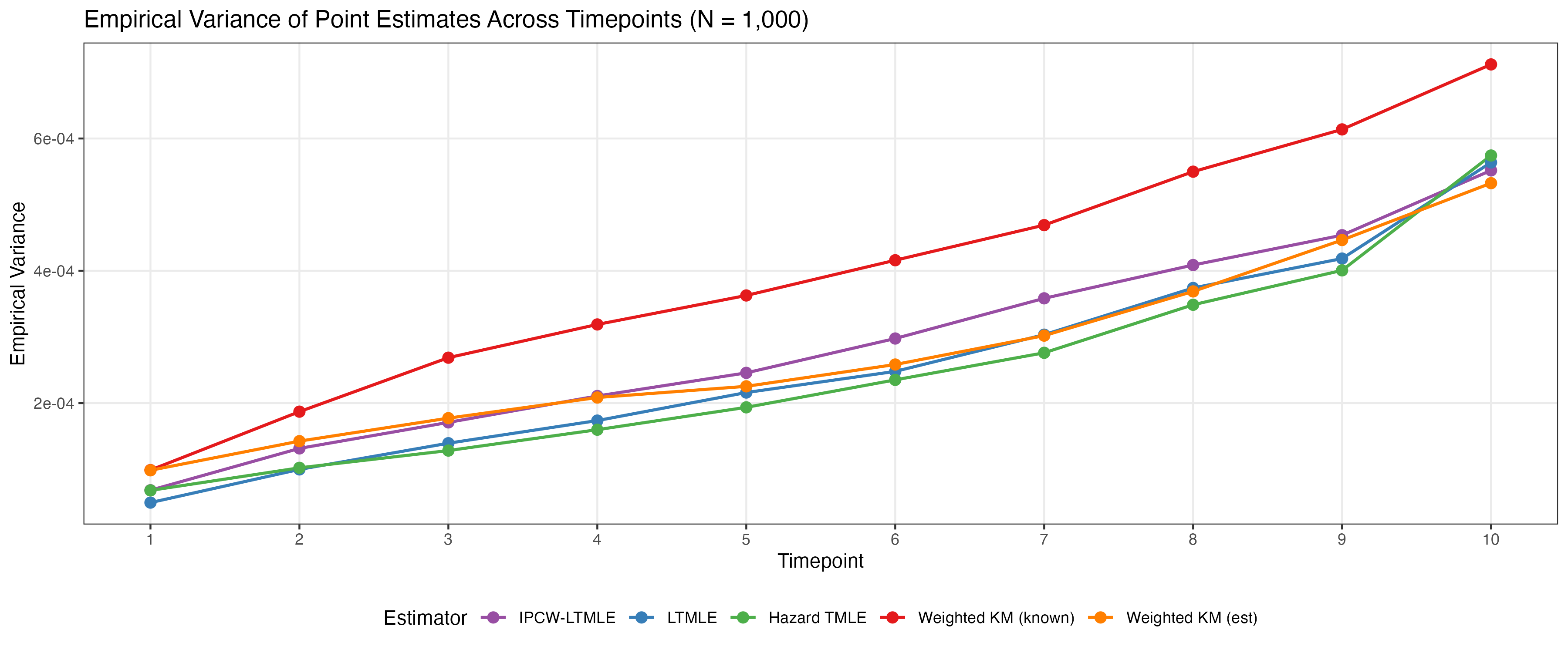}  
\caption{Variance of the point estimate for each estimator across 1,000 simulations of sample size N=1,000.}
  \label{fig:N1k_var_graph}
\end{figure}

\begin{sidewaystable}
\centering
\caption{Results from 1k simulation studies with sample size $N = 1$k per simulation. Variance reported is the variance of the point estimate across simulations. Inference for each estimator which supports 95\% confidence interval coverage is described in Section~\ref{subsec:estimator_comp}. Here, (est) refers to estimated $\Pi_{\Delta}$ and (known) refers to the use of known values of $\Pi_{\Delta}$.} 
\label{tab:results_N1000}
\begin{tabular}{llllllllllll}
  \toprule
Metric & Estimator & $t=1$ & $t=2$ & $t=3$ & $t=4$ & $t=5$ & $t=6$ & $t=7$ & $t=8$ & $t=9$ & $t=10$ \\ 
  \midrule
bias & IPCW-LTMLE & -0.000861 & -0.000315 & -0.000483 & -0.000573 & -0.000462 & -0.000581 & -0.000538 & -0.000507 & -0.000022 & 0.000706 \\ 
   & LTMLE & -0.001524 & 0.000231 & 0.000466 & 0.000283 & -0.000431 & -0.000442 & -0.001004 & -0.000180 & -0.000575 & -0.000283 \\ 
   & Hazard TMLE & -0.000104 & -0.000359 & -0.000000 & -0.000141 & -0.000209 & 0.000029 & -0.000049 & 0.000866 & 0.000394 & 0.000669 \\ 
   & Weighted KM (known) & 0.000068 & 0.000140 & 0.000281 & 0.000531 & 0.000703 & 0.000472 & 0.001039 & 0.001050 & 0.001564 & 0.001547 \\ 
   & Weighted KM (est) & -0.000301 & -0.000073 & 0.000418 & 0.000065 & 0.000999 & 0.001287 & 0.001417 & 0.000449 & 0.000196 & 0.000478 \\ 
   \midrule
var & IPCW-LTMLE & 0.000068 & 0.000132 & 0.000171 & 0.000211 & 0.000246 & 0.000298 & 0.000358 & 0.000409 & 0.000454 & 0.000552 \\ 
   & LTMLE & 0.000050 & 0.000100 & 0.000139 & 0.000174 & 0.000216 & 0.000248 & 0.000303 & 0.000374 & 0.000418 & 0.000564 \\ 
   & Hazard TMLE & 0.000068 & 0.000102 & 0.000128 & 0.000160 & 0.000194 & 0.000235 & 0.000276 & 0.000349 & 0.000401 & 0.000574 \\ 
   & Weighted KM (known) & 0.000099 & 0.000187 & 0.000269 & 0.000319 & 0.000363 & 0.000416 & 0.000469 & 0.000550 & 0.000614 & 0.000712 \\ 
   & Weighted KM (est) & 0.000098 & 0.000143 & 0.000177 & 0.000208 & 0.000225 & 0.000258 & 0.000302 & 0.000369 & 0.000446 & 0.000532 \\ 
   \midrule
mse & IPCW-LTMLE & 0.000069 & 0.000131 & 0.000171 & 0.000211 & 0.000246 & 0.000298 & 0.000358 & 0.000409 & 0.000453 & 0.000552 \\ 
   & LTMLE & 0.000052 & 0.000100 & 0.000139 & 0.000173 & 0.000216 & 0.000248 & 0.000304 & 0.000374 & 0.000418 & 0.000563 \\ 
   & Hazard TMLE & 0.000068 & 0.000102 & 0.000128 & 0.000160 & 0.000193 & 0.000235 & 0.000276 & 0.000349 & 0.000400 & 0.000574 \\ 
   & Weighted KM (known) & 0.000099 & 0.000187 & 0.000268 & 0.000319 & 0.000363 & 0.000416 & 0.000470 & 0.000550 & 0.000616 & 0.000714 \\ 
   & Weighted KM (est) & 0.000098 & 0.000142 & 0.000177 & 0.000208 & 0.000226 & 0.000260 & 0.000304 & 0.000369 & 0.000446 & 0.000532 \\ 
   \midrule
coverage & IPCW-LTMLE & 0.941 & 0.925 & 0.936 & 0.935 & 0.933 & 0.935 & 0.940 & 0.934 & 0.941 & 0.950 \\ 
   & LTMLE & 0.963 & 0.930 & 0.933 & 0.939 & 0.937 & 0.934 & 0.940 & 0.938 & 0.952 & 0.949 \\ 
   & Hazard TMLE & 0.922 & 0.933 & 0.937 & 0.937 & 0.953 & 0.936 & 0.947 & 0.950 & 0.954 & 0.968 \\ 
   & Weighted KM (known) & 0.914 & 0.922 & 0.931 & 0.933 & 0.943 & 0.945 & 0.950 & 0.947 & 0.933 & 0.940 \\ 
   & Weighted KM (est) & 0.908 & 0.945 & 0.941 & 0.947 & 0.942 & 0.941 & 0.942 & 0.937 & 0.941 & 0.941 \\ 
   \bottomrule
\end{tabular}
\end{sidewaystable}

\newpage 

\subsubsection{Sample size N=3,000}
\begin{figure}[H]
  \centering
  \includegraphics[width=\textwidth]{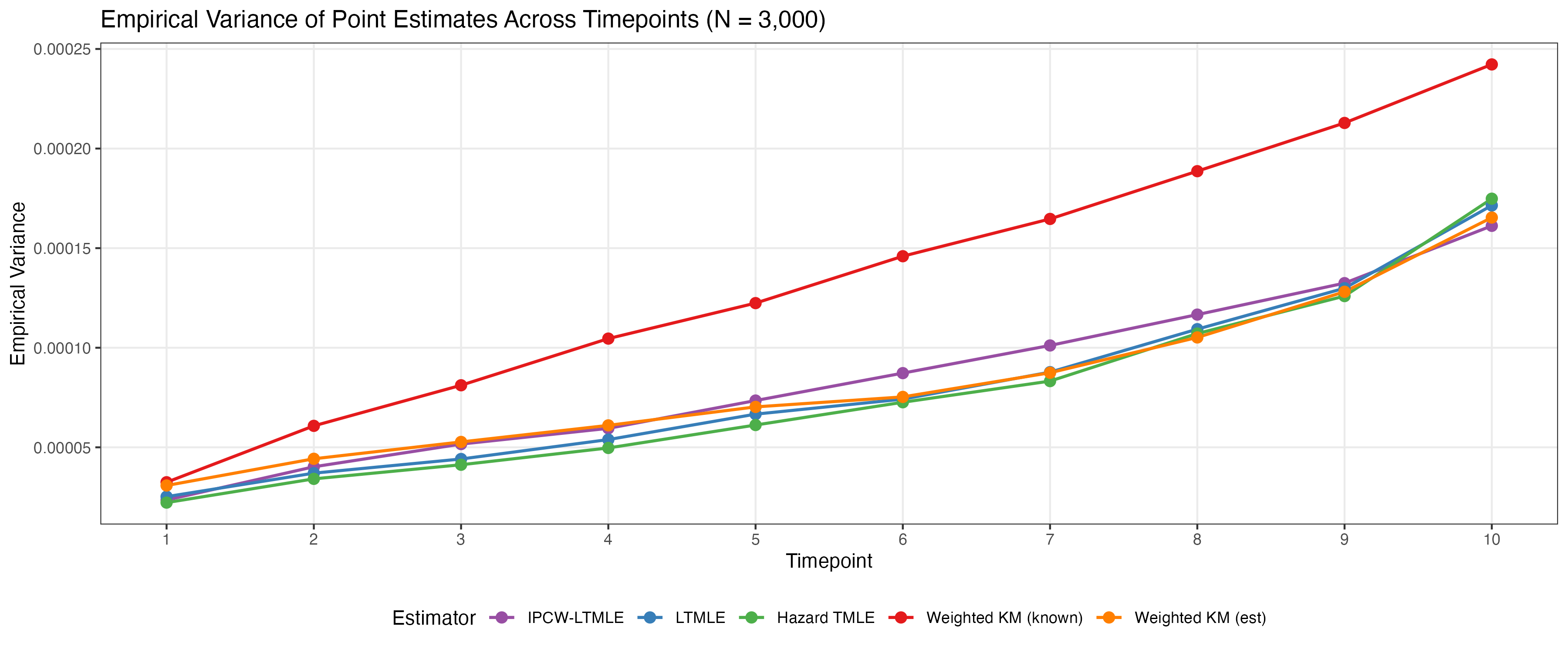}  
\caption{Variance of the point estimate for each estimator across 1,000 simulations of sample size N=3,000.}
  \label{fig:N3k_graph}
\end{figure}

\begin{sidewaystable}
\centering
\caption{Results from 1k simulation studies with sample size $N = 3$k per simulation. Variance reported is the variance of the point estimate across simulations. Inference for each estimator which supports 95\% confidence interval coverage is described in Section~\ref{subsec:estimator_comp}. Here, (est) refers to estimated $\Pi_{\Delta}$ and (known) refers to the use of known values of $\Pi_{\Delta}$.} 
\label{tab:results_N3000}
\begin{tabular}{llllllllllll}
  \toprule
Metric & Estimator & $t=1$ & $t=2$ & $t=3$ & $t=4$ & $t=5$ & $t=6$ & $t=7$ & $t=8$ & $t=9$ & $t=10$ \\ 
  \midrule
bias & IPCW-LTMLE & -0.000081 & -0.000075 & -0.000129 & -0.000146 & -0.000174 & 0.000212 & 0.000066 & -0.000140 & -0.000039 & 0.000080 \\ 
   & LTMLE & -0.000053 & -0.000231 & -0.000178 & -0.000001 & -0.000197 & 0.000280 & 0.000097 & 0.000046 & -0.000054 & -0.000779 \\ 
   & Hazard TMLE & -0.000125 & -0.000169 & -0.000119 & 0.000085 & -0.000009 & 0.000434 & 0.000324 & 0.000291 & 0.000227 & -0.000316 \\ 
   & Weighted KM (known) & -0.000363 & -0.000516 & -0.000045 & -0.000228 & -0.000429 & -0.000439 & -0.000619 & -0.000380 & -0.000101 & -0.000059 \\ 
   & Weighted KM (est) & 0.000125 & 0.000187 & 0.000583 & 0.000349 & 0.000352 & 0.000381 & 0.000403 & 0.000143 & 0.000103 & 0.000428 \\ 
   \midrule
var & IPCW-LTMLE & 0.000024 & 0.000040 & 0.000052 & 0.000060 & 0.000073 & 0.000087 & 0.000101 & 0.000117 & 0.000132 & 0.000161 \\ 
   & LTMLE & 0.000025 & 0.000037 & 0.000044 & 0.000054 & 0.000067 & 0.000074 & 0.000088 & 0.000109 & 0.000130 & 0.000171 \\ 
   & Hazard TMLE & 0.000022 & 0.000034 & 0.000041 & 0.000050 & 0.000061 & 0.000073 & 0.000083 & 0.000107 & 0.000126 & 0.000175 \\ 
   & Weighted KM (known) & 0.000032 & 0.000061 & 0.000081 & 0.000105 & 0.000122 & 0.000146 & 0.000165 & 0.000189 & 0.000213 & 0.000242 \\ 
   & Weighted KM (est) & 0.000031 & 0.000044 & 0.000053 & 0.000061 & 0.000070 & 0.000075 & 0.000087 & 0.000105 & 0.000128 & 0.000165 \\ 
   \midrule
mse & IPCW-LTMLE & 0.000024 & 0.000040 & 0.000052 & 0.000059 & 0.000073 & 0.000087 & 0.000101 & 0.000117 & 0.000132 & 0.000161 \\ 
   & LTMLE & 0.000025 & 0.000037 & 0.000044 & 0.000054 & 0.000067 & 0.000074 & 0.000088 & 0.000109 & 0.000130 & 0.000172 \\ 
   & Hazard TMLE & 0.000022 & 0.000034 & 0.000041 & 0.000050 & 0.000061 & 0.000073 & 0.000083 & 0.000107 & 0.000126 & 0.000175 \\ 
   & Weighted KM (known) & 0.000033 & 0.000061 & 0.000081 & 0.000105 & 0.000122 & 0.000146 & 0.000165 & 0.000189 & 0.000213 & 0.000242 \\ 
   & Weighted KM (est) & 0.000031 & 0.000044 & 0.000053 & 0.000061 & 0.000070 & 0.000075 & 0.000088 & 0.000105 & 0.000128 & 0.000165 \\ 
   \midrule
coverage & IPCW-LTMLE & 0.936 & 0.941 & 0.955 & 0.949 & 0.945 & 0.961 & 0.953 & 0.952 & 0.958 & 0.956 \\ 
   & LTMLE & 0.930 & 0.941 & 0.943 & 0.955 & 0.945 & 0.950 & 0.955 & 0.946 & 0.948 & 0.951 \\ 
   & Hazard TMLE & 0.938 & 0.944 & 0.948 & 0.946 & 0.943 & 0.942 & 0.953 & 0.941 & 0.951 & 0.948 \\ 
   & Weighted KM (known) & 0.938 & 0.946 & 0.951 & 0.934 & 0.946 & 0.939 & 0.936 & 0.939 & 0.939 & 0.942 \\ 
   & Weighted KM (est) & 0.924 & 0.937 & 0.953 & 0.937 & 0.941 & 0.949 & 0.945 & 0.956 & 0.946 & 0.942 \\ 
   \bottomrule
\end{tabular}
\end{sidewaystable}

\newpage 

\subsubsection{Variance estimation}
\label{subsubsec:variance_sims}

In this section, we present plots of coverage for each estimator's different variance estimation procedures. For the LTMLE and hazard TMLE, we present cross-fit versus non-cross-fitted variance estimators. For the IPCW-LTMLE we present the four variance estimators as described previously -- representing combinations of cross-fitting versus not cross-fitting and using vs not using the projection correction. Tables with more detailed simulation results including mean standard error, average 95\% confidence interval coverage, and ratio of the estimated versus true standard error for all estimators with and without cross-fitting can be found in the  Supplementary Material~\ref{sec:supp}.

\begin{figure}[H]
  \centering
  \includegraphics[width=\textwidth]{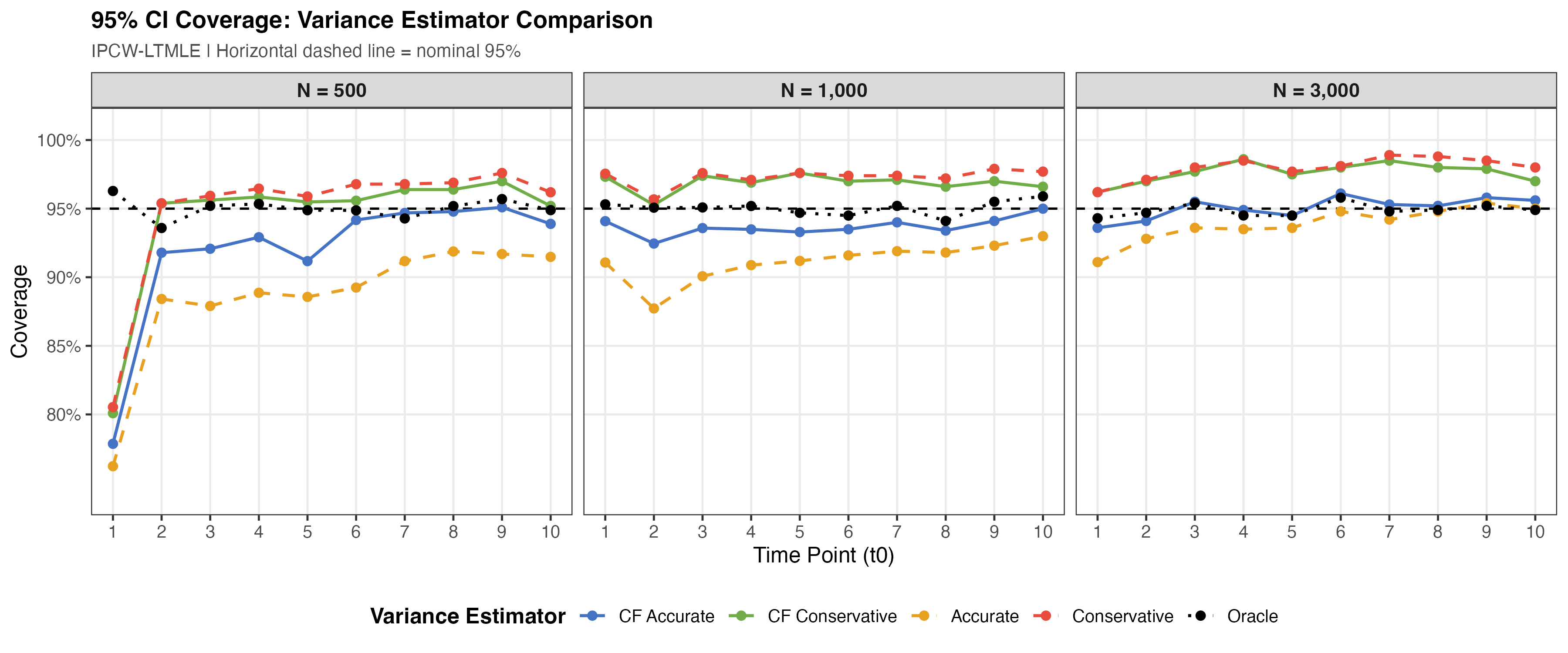}  
\caption{Confidence interval coverage for targeted IPCW-LTMLE across 1,000 simulations of sample size $N \in \{500, 1000, 3000\}$.}
  \label{fig:ipcwltmle_cov}
\end{figure}

\begin{figure}[H]
  \centering
  \includegraphics[width=\textwidth]{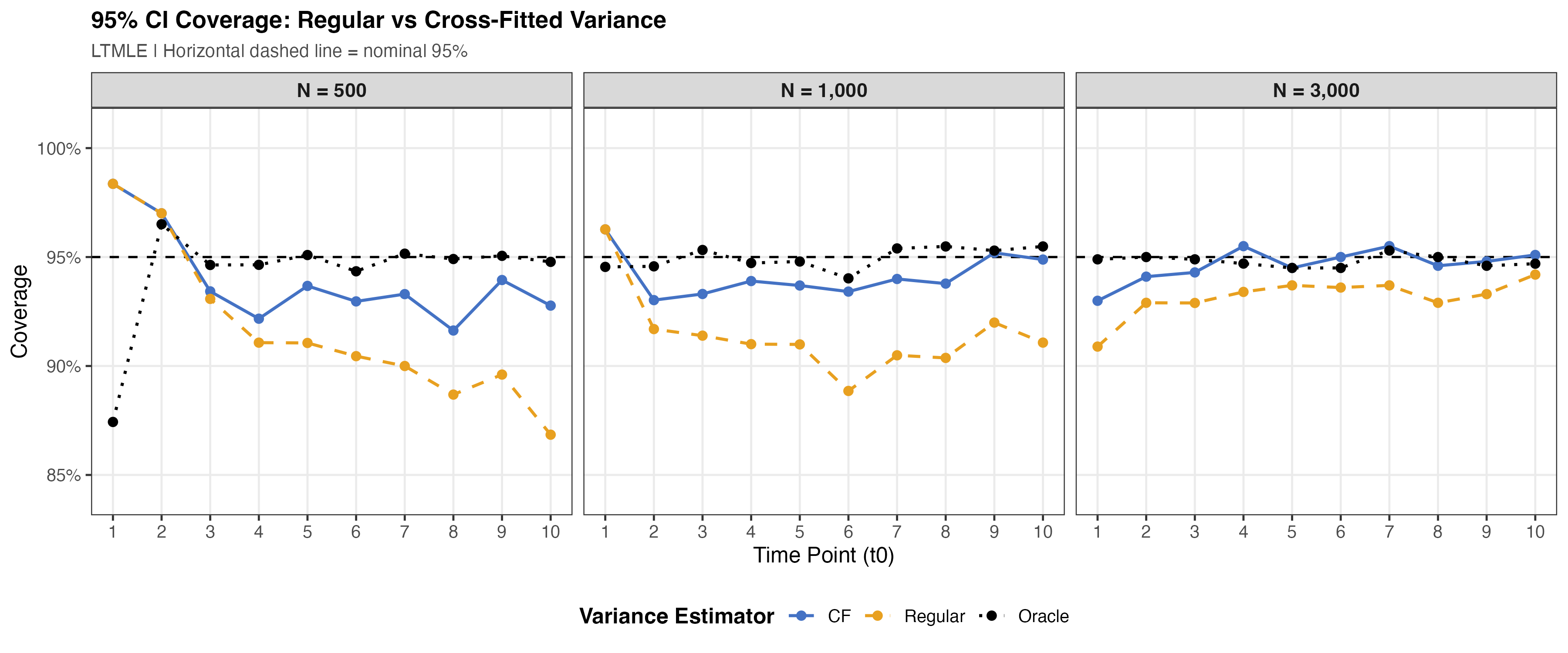}  
\caption{Confidence interval coverage for targeted LTMLE across 1,000 simulations of sample size $N \in \{500, 1000, 3000\}$.}
  \label{fig:ltmle_cov}
\end{figure}

\begin{figure}[H]
  \centering
  \includegraphics[width=\textwidth]{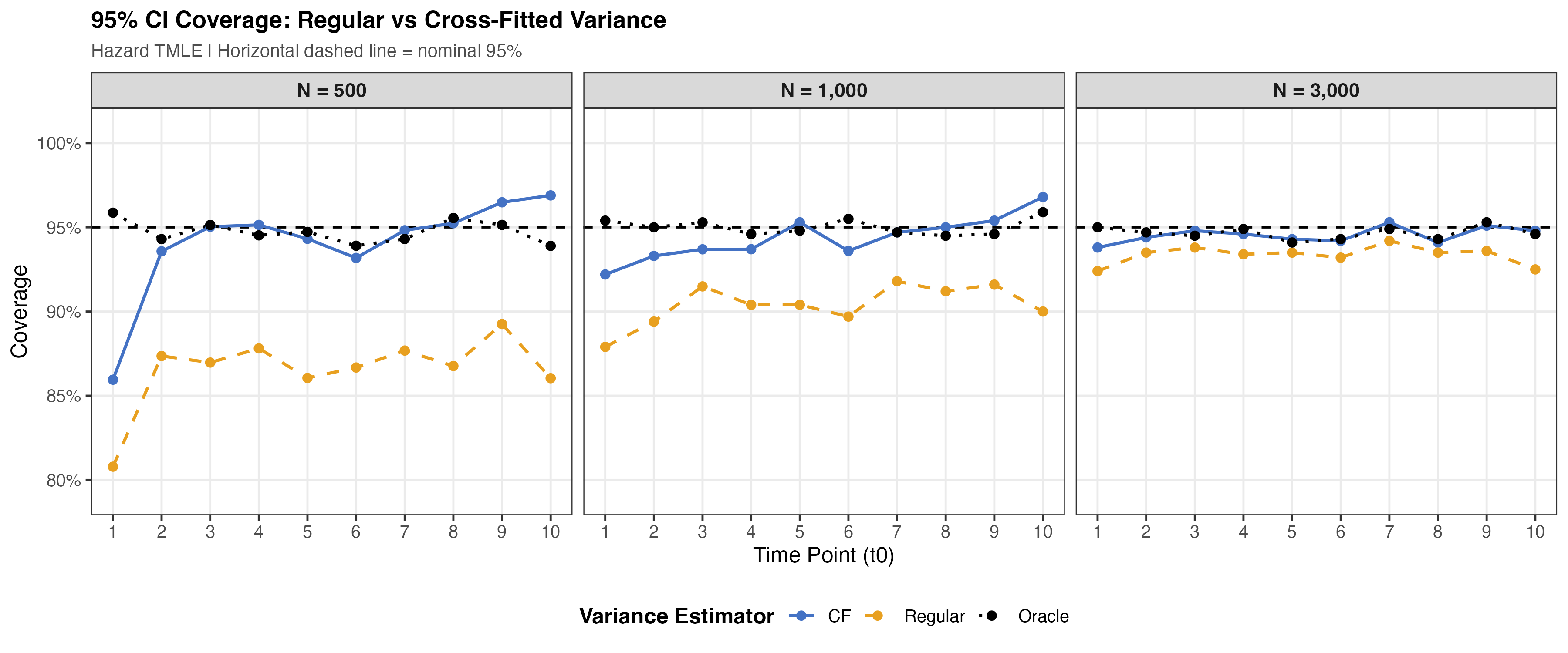}  
\caption{Confidence interval coverage for hazard TMLE across 1,000 simulations of sample size $N \in \{500, 1000, 3000\}$.}
  \label{fig:haz_cov}
\end{figure}

\section{Discussion}
\label{sec:disc}

This paper began with a practical problem: existing estimators for 
causal survival parameters in resampling designs are inefficient, 
discarding the rich longitudinal covariate history available in these 
studies and relying on inverse probability weights that are known by 
design but suboptimal from an efficiency perspective. In formalizing the 
resampling data structure, we recognized that resampling designs are 
an instance of a broader class, two-stage designs with outcome 
subsampling, and that this connection opens the door to more 
efficient estimation strategies. Drawing on the IPCW-TMLE framework of 
\citet{rose2011targeted}, we developed an IPCW-LTMLE for the longitudinal 
setting and showed that targeting the known sampling weights yields 
meaningful efficiency gains. Recognizing that inverse weighting itself sacrifices 
efficiency regardless of how carefully the weights are handled, we 
then proposed a novel LTMLE that incorporates the second-stage sampling 
mechanism directly as an intervention node, returning to plug-in estimation 
principles. Finally, initial simulation studies revealed that standard 
variance estimation fails badly in both estimators when flexible nuisance 
functions are used, motivating the development of cross-fitted variance 
estimators that restore nominal coverage.

As established in \citet{landsiedel2025hazard} and reiterated here, 
two-stage designs with administrative or right censoring and outcome 
missingness due to the second-stage sampling $\Delta$ fall into the class of 
bivariate censored data problems. For such problems, fully efficient 
closed-form estimators do not exist in general \citep{van1996efficient, quale2002locally}. The estimators proposed in 
this paper are therefore 
best understood as highly efficient closed-form alternatives that add to the 
growing toolbox of estimators for resampling and two-stage designs. Together 
with our prior work, these estimators represent a systematic effort to bring 
modern semiparametric efficiency theory to bear on a class of problems that 
has historically relied on simple inverse-weighting approaches.

A central motivation for this work was the observation that the weighted 
Kaplan-Meier estimator with known weights, despite its widespread use in two-stage designs studying
HIV-related mortality \citep{yiannoutsos2008sampling, geng2015estimation, 
holmes2018estimated}, discards all information from participants with 
$\Delta = 0$ and ignores the rich longitudinal covariate history available 
in these studies. Our simulation results confirm that this inefficiency is 
practically meaningful, especially at early time points. This insight applies to inverse weighted estimators across all two-stage designs; weights are typically known by design, yet they are not optimal from an efficiency perspective.

The simulation results reveal a clear hierarchy among the estimators, with 
meaningful structure in the sources of efficiency gains and losses. The full 
LTMLE achieves the lowest empirical variance across most settings, consistent 
with its status as a full plug-in estimator that avoids inverse probability 
weighting entirely and leverages the 
longitudinal covariate history of all participants. Hazard-TMLE is similarly competitive through most of the follow-up 
period: while it involves inverse weighting by administrative censoring, which 
becomes relevant after $t = 5$ in our simulations, it avoids direct inverse 
weighting by $\Delta$ and benefits from some of the same plug-in principles. IPCW-LTMLE, despite targeting the resampling weights to partially compensate 
for lost efficiency, cannot fully recover the variance advantage of the full 
LTMLE; the immediate loss of efficiency due to inverse weighting by $\Delta$ is not offset even with careful estimation and weight targeting. Among the 
non-TMLE estimators, weighted Kaplan-Meier with known weights performs worst 
overall, as it discards all information from $\Delta = 0$ participants, 
including those known to be alive and uncensored through a substantial portion 
of follow-up before becoming LTFU. Weighted 
Kaplan-Meier with estimated weights is more competitive, particularly at later 
timepoints where it can be more efficient than the TMLEs; by avoiding an outcome model entirely, it sidesteps the challenge 
of fitting outcome regressions on sparse survival data near the end of 
follow-up, which may explain its relatively stable tail performance at $t = 10$. 
However, this comes at the cost of efficiency at earlier timepoints, where the 
rich longitudinal covariate history provides substantial leverage that an 
outcome-model-free estimator cannot exploit.  For example, an individual who 
is observed at $t = 9$ but lost to follow-up before $t = 10$ receives zero 
weight under wKM and contributes nothing to the estimator; the TMLE estimators 
instead impute the counterfactual outcome at $t = 10$ using the observed 
covariate history, which is highly predictive of survival status. This 
efficiency gain cannot be fully recovered from recovered by incorporating covariate information into 
the weight estimation step of wKM.

The sampling distributions of the point estimates at $t = 1$ and $n = 500$ 
(Section~\ref{sec:supp}) provide further insight into the finite-sample 
behavior of the coverage of these estimators in another challenging region (recall all estimators had over or undercoverage at $t=1$ with $n=500$). The hazard-TMLE 
sampling distribution at $t=1$ is right-skewed, while that of IPCW-LTMLE is notably 
bimodal, a pattern consistent with instability induced by inverse weighting 
in a region where near-deterministic covariate histories and a rare outcome (few deaths) produce extreme weights 
for a small number of observations. In both 
cases, departure from normality undermines the Wald-type confidence intervals. The LTMLE sampling distribution 
is also skewed at $t = 1$, consistent with its oracle undercoverage in this 
region; however, as a full plug-in estimator that avoids inverse weighting by 
$\Delta$, it does not exhibit the same bimodality. These $t = 1$ pathologies 
are a finite-sample phenomenon: coverage is near-nominal for all estimators at 
$n = 1{,}000$ and $n = 3{,}000$. Taken together, these results suggest that 
the $t = 1$ region at small sample sizes presents a fundamentally difficult 
estimation problem in which low event counts, vast deterministic information, 
and departures from normality challenge all three estimators in different ways.

Several limitations merit discussion. First, our simulations were conducted 
under a single data generating process, most similar to that which arises in resampling based studies of HIV-related mortality. Second, the computational demands of the proposed estimators are non-trivial. Our TMLE-based estimators all nest cross-validated machine learning within cross-fit procedures for variance estimation. While these strategies are well-intended attempts at reducing bias and maintaining proper confidence interval coverage, the computational time adds up. Third, in our simulations, our observed data come from a bivariate censored version of a full data process, so we are not able to construct fully efficient estimators in closed form. The estimators we present in this work represent closed-form alternatives that we believe to be highly efficient. However, future work might include attempting to construct a fully efficient estimator in this setting -- if for no other reason than to assess how much efficiency is lost by using such closed-form estimators. Last, we did not address how to proceed with estimation when second-stage data collection is attempted but fails for a given subset of participants; we leave this important consideration for future work. 

This paper provides a foundation for more efficient and robust analysis of general
causal parameters in longitudinal two-stage designs with outcome subsampling. By formalizing the connection between resampling and two-stage designs, 
proposing two new estimators for general causal parameters, and demonstrating the necessity of cross-fitted variance 
estimation, we add to the growing body of methods around these unique study designs. The 
resulting estimators are more efficient, more robust, and, with 
cross-fitted variance, more reliable than existing approaches, offering a 
principled path forward for causal analyses in two-stage designs with outcome subsampling.

\newpage
\bibliography{references}
\bibliographystyle{apalike}

\clearpage
\appendix
\section{Supplementary material}
\label{sec:supp}

\subsection{Table of results for cross-fitted variance estimation}

\subsubsection{IPCW-LTMLE, $N=500$}

\begin{figure}[H]
  \centering
  \includegraphics[width=\textwidth]{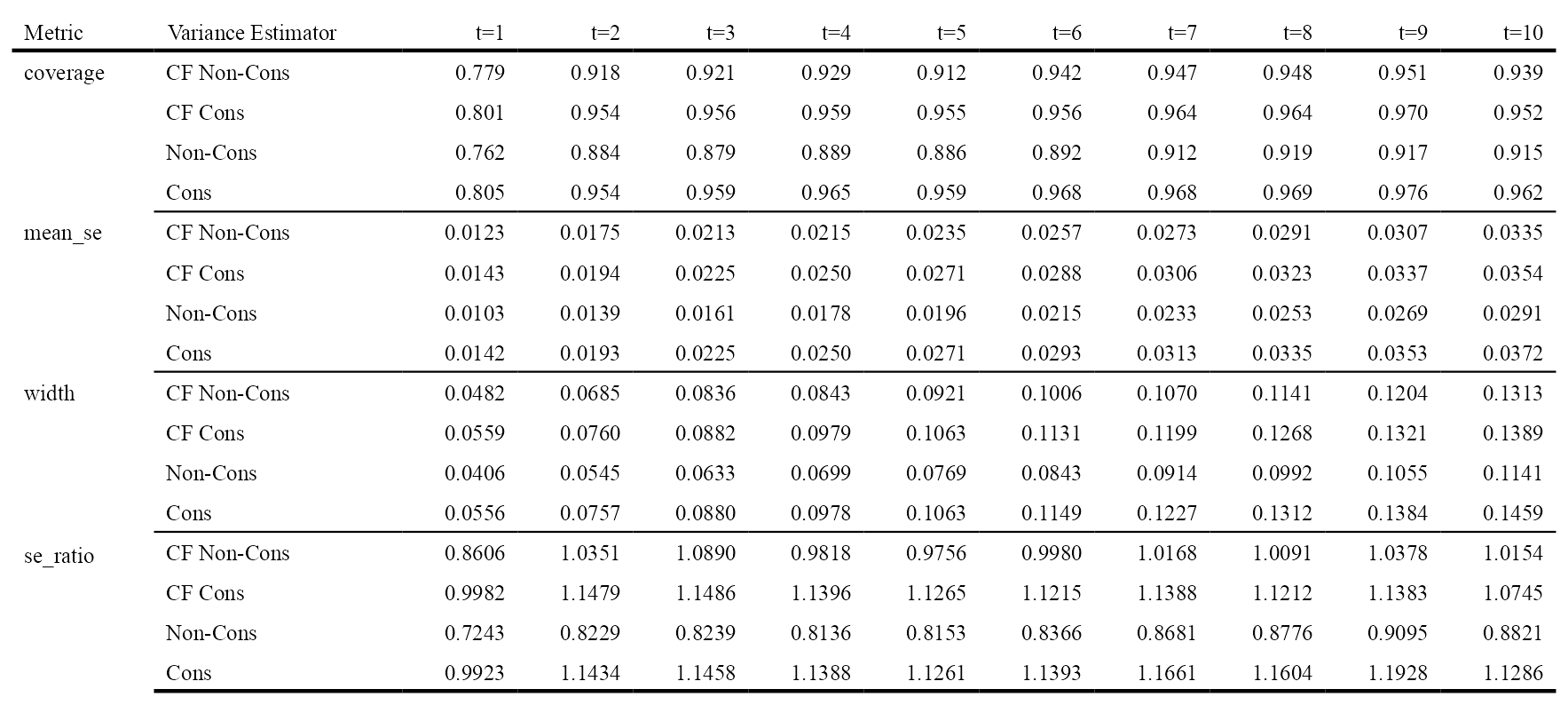}
  \caption{Coverage, mean standard error (mean\_se), 95\% confidence interval width (width), and standard error ratio (se\_ratio) relative to the true (empirical se) for the four variance estimators for IPCW-LTMLE at sample size $N=500$.}
  \label{fig:N500_cross_ipcw}
\end{figure}

\subsubsection{IPCW-LTMLE, $N=1{,}000$}

\begin{figure}[H]
  \centering
  \includegraphics[width=\textwidth]{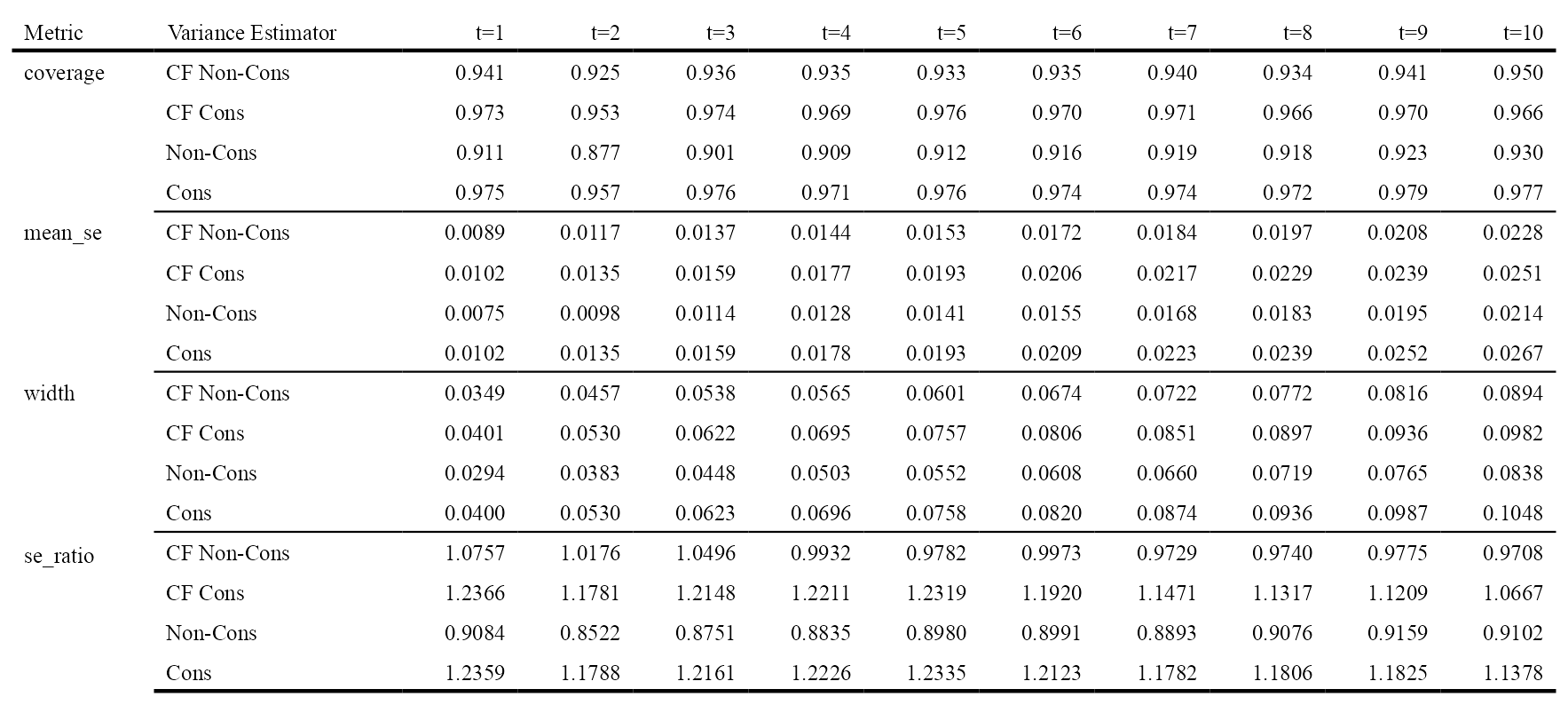}
  \caption{Coverage, mean standard error (mean\_se), 95\% confidence interval width (width), and standard error ratio (se\_ratio) relative to the true (empirical se) for the four variance estimators for IPCW-LTMLE at sample size $N=1{,}000$.}
  \label{fig:N1k_cross_ipcw}
\end{figure}

\subsubsection{IPCW-LTMLE, $N=3{,}000$}

\begin{figure}[H]
  \centering
  \includegraphics[width=\textwidth]{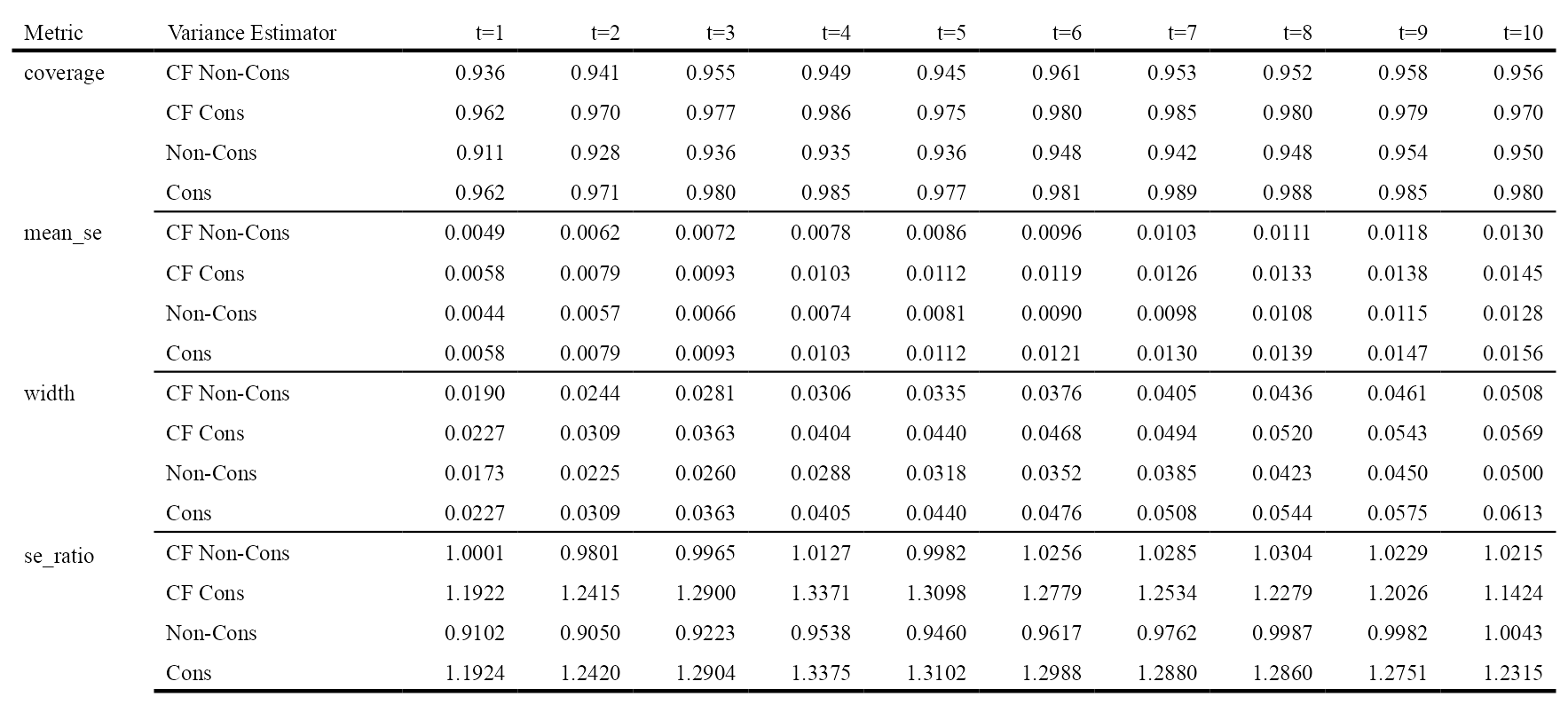}
  \caption{Coverage, mean standard error (mean\_se), 95\% confidence interval width (width), and standard error ratio (se\_ratio) relative to the true (empirical se) for the four variance estimators for IPCW-LTMLE at sample size $N=3{,}000$.}
  \label{fig:N3k_cross_ipcw}
\end{figure}

\subsubsection{LTMLE, $N=500$}

\begin{figure}[H]
  \centering
  \includegraphics[width=\textwidth]{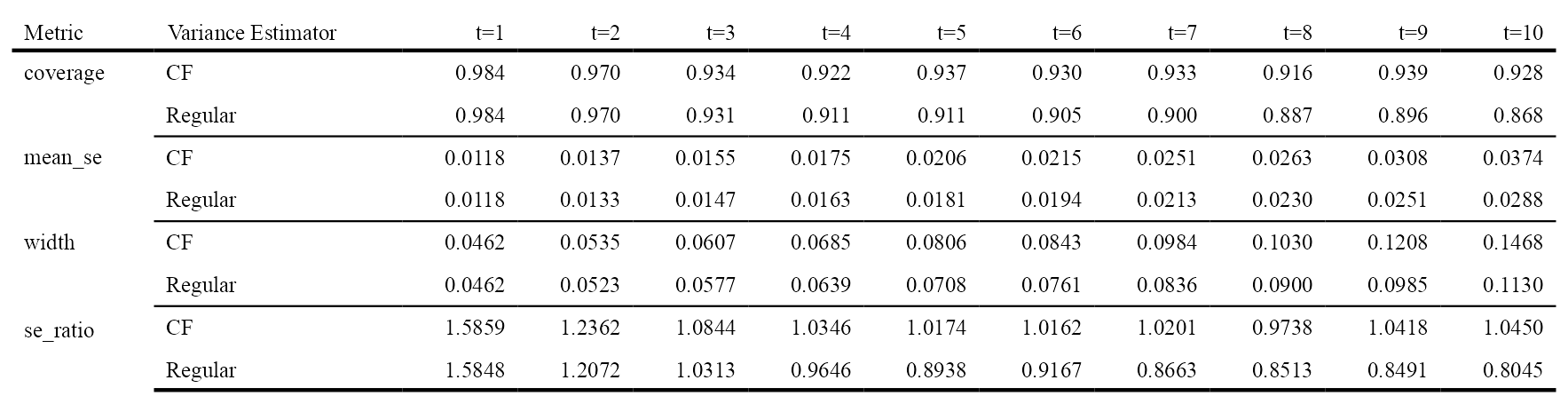}
  \caption{Coverage, mean standard error (mean\_se), 95\% confidence interval width (width), and standard error ratio (se\_ratio) relative to the true (empirical se) for the regular (non-cross-fit) versus cross-fit (CF) variance estimators for LTMLE at sample size $N=500$.}
  \label{fig:N500_cross_ltmle}
\end{figure}

\subsubsection{LTMLE, $N=1{,}000$}

\begin{figure}[H]
  \centering
  \includegraphics[width=\textwidth]{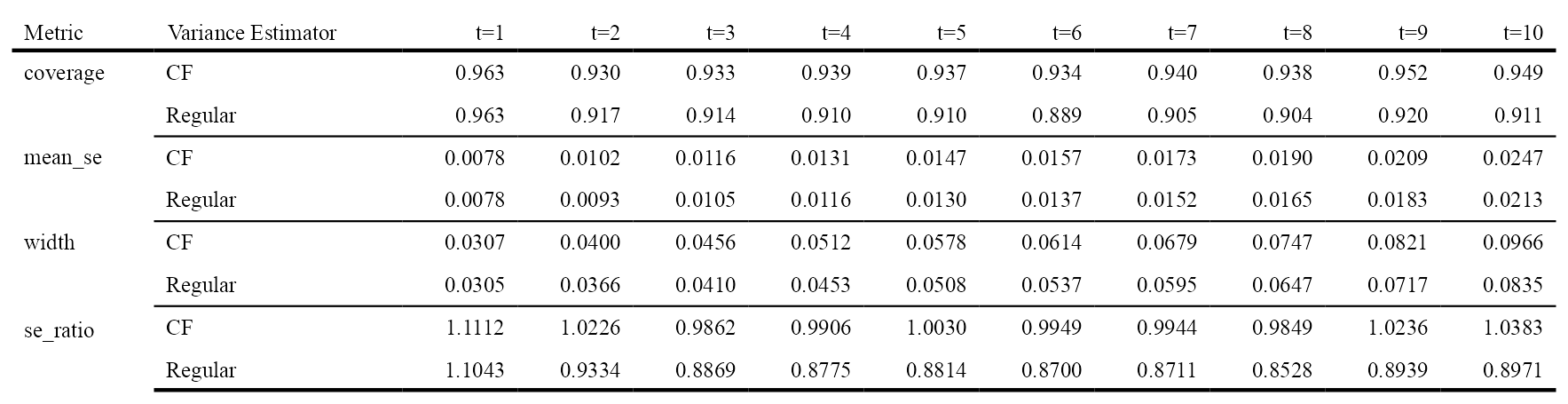}
  \caption{Coverage, mean standard error (mean\_se), 95\% confidence interval width (width), and standard error ratio (se\_ratio) relative to the true (empirical se) for the regular (non-cross-fit) versus cross-fit (CF) variance estimators for LTMLE at sample size $N=1{,}000$.}
  \label{fig:N1k_cross_ltmle}
\end{figure}

\subsubsection{LTMLE, $N=3{,}000$}

\begin{figure}[H]
  \centering
  \includegraphics[width=\textwidth]{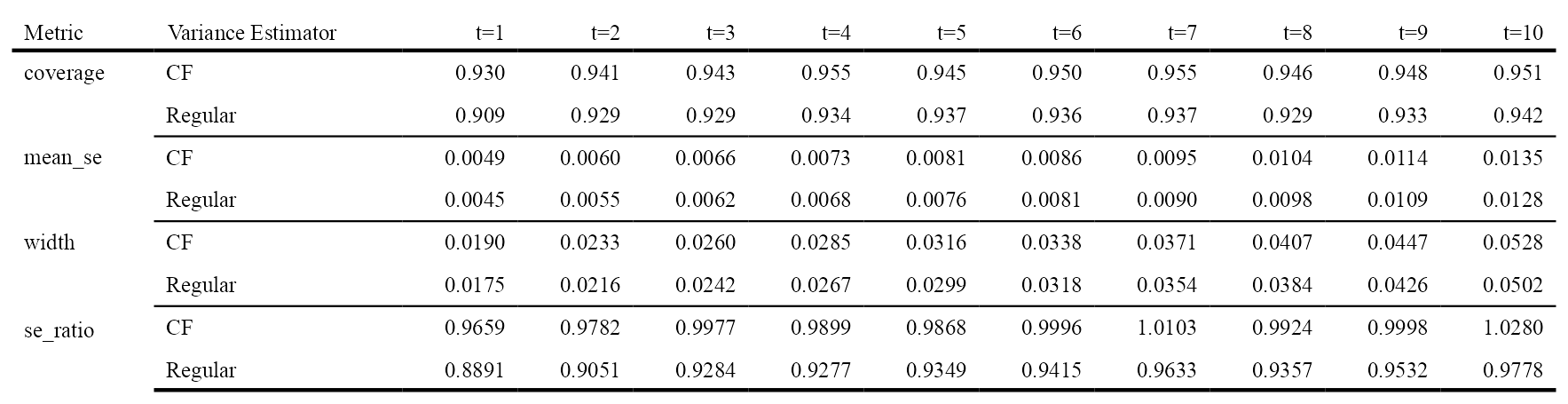}
  \caption{Coverage, mean standard error (mean\_se), 95\% confidence interval width (width), and standard error ratio (se\_ratio) relative to the true (empirical se) for the regular (non-cross-fit) versus cross-fit (CF) variance estimators for LTMLE at sample size $N=3{,}000$.}
  \label{fig:N3k_cross_ltmle}
\end{figure}

\subsubsection{Hazard TMLE, $N=500$}

\begin{figure}[H]
  \centering
  \includegraphics[width=\textwidth]{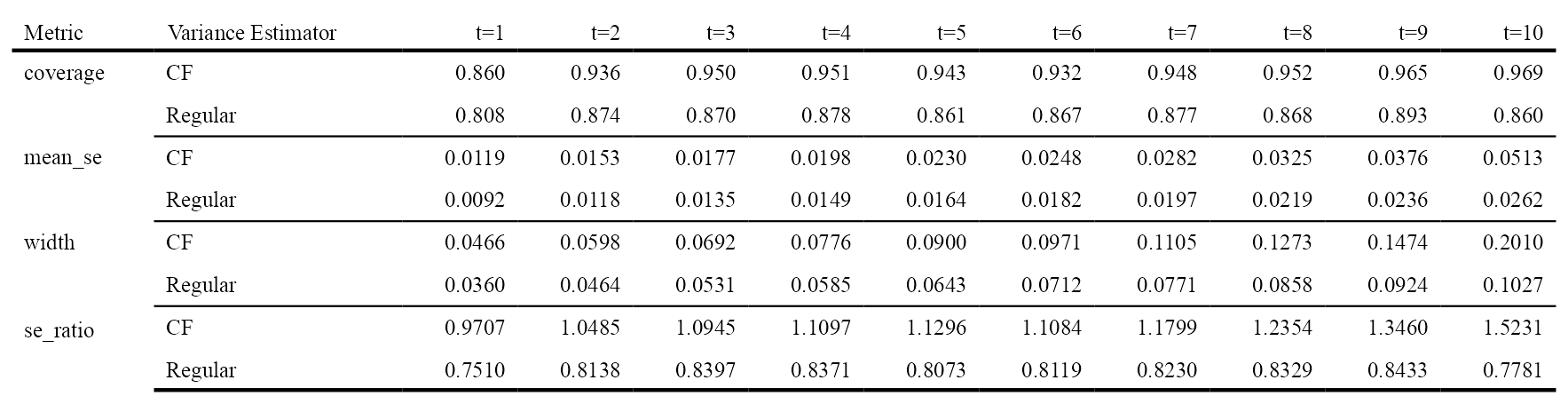}
  \caption{Coverage, mean standard error (mean\_se), 95\% confidence interval width (width), and standard error ratio (se\_ratio) relative to the true (empirical se) for the regular (non-cross-fit) versus cross-fit (CF) variance estimators for Hazard TMLE at sample size $N=500$.}
  \label{fig:N500_cross_haz}
\end{figure}

\subsubsection{Hazard TMLE, $N=1{,}000$}

\begin{figure}[H]
  \centering
  \includegraphics[width=\textwidth]{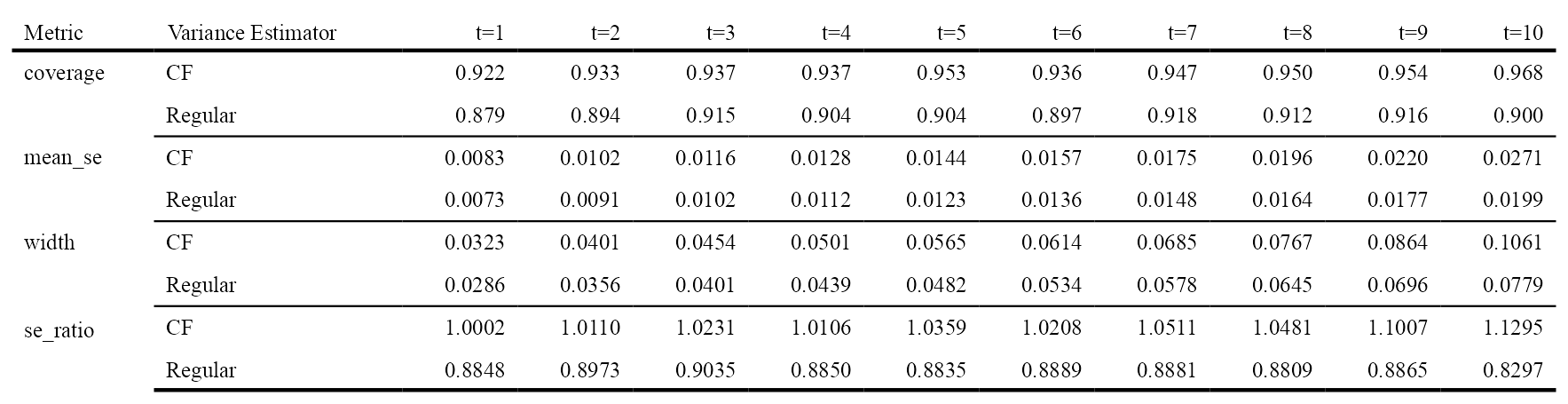}
  \caption{Coverage, mean standard error (mean\_se), 95\% confidence interval width (width), and standard error ratio (se\_ratio) relative to the true (empirical se) for the regular (non-cross-fit) versus cross-fit (CF) variance estimators for Hazard TMLE at sample size $N=1{,}000$.}
  \label{fig:N1k_cross_haz}
\end{figure}

\subsubsection{Hazard TMLE, $N=3{,}000$}

\begin{figure}[H]
  \centering
  \includegraphics[width=\textwidth]{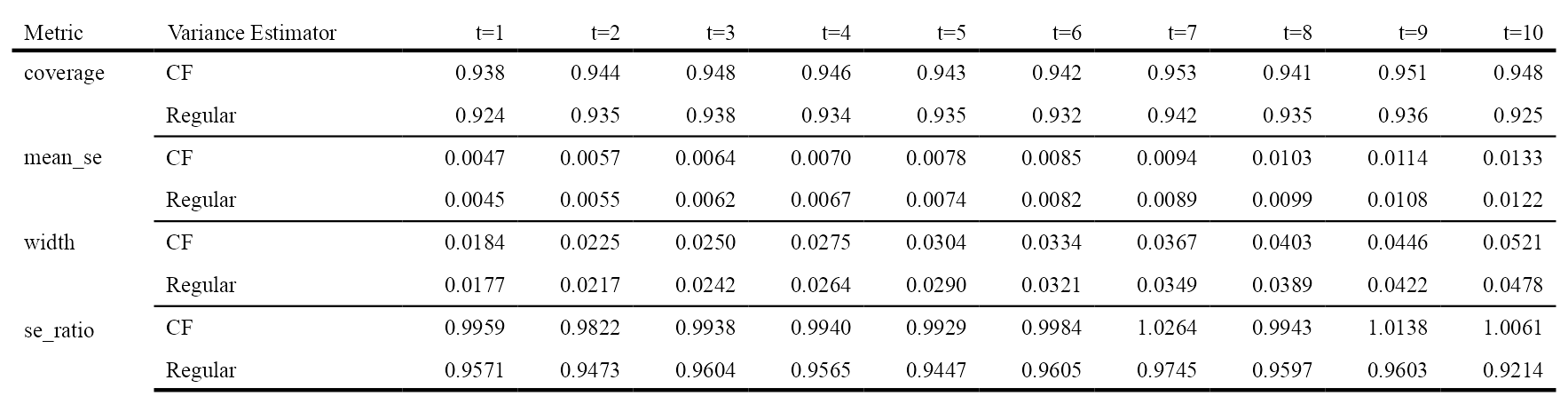}
  \caption{Coverage, mean standard error (mean\_se), 95\% confidence interval width (width), and standard error ratio (se\_ratio) relative to the true (empirical se) for the regular (non-cross-fit) versus cross-fit (CF) variance estimators for Hazard TMLE at sample size $N=3{,}000$.}
  \label{fig:N3k_cross_haz}
\end{figure}

\subsection{Distribution of point estimates, $N=500$, $t=1$}

\begin{figure}[H]
  \centering
  \includegraphics[width=\textwidth]{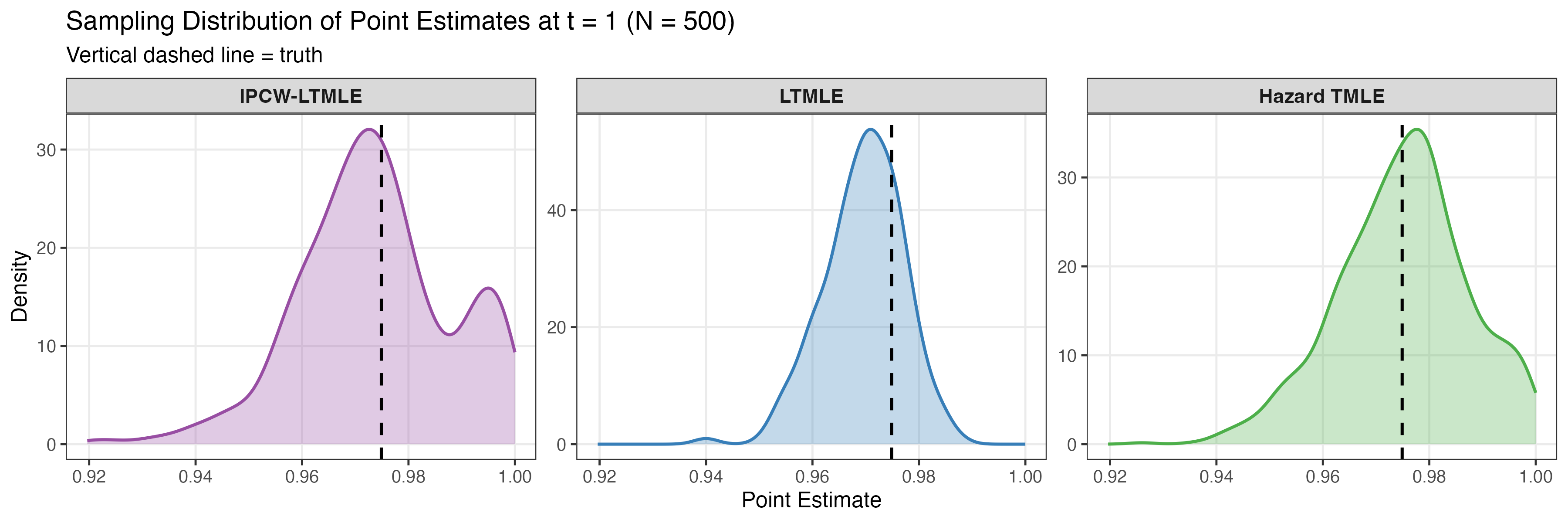}
  \caption{Sampling distribution of the point estimate for IPCW-LTMLE, LTMLE, and Hazard TMLE at $t=1$, $N=500$.}
  \label{fig:point_est}
\end{figure}

\end{document}